\documentclass[%
preprintnumbers,
 amsmath,amssymb,
 aps,
]{revtex4-2}

\usepackage{graphicx}
\usepackage{dcolumn}
\usepackage{bm}

\usepackage{subcaption, float}
\usepackage{epstopdf}
\newcommand{\pder}[2]{\frac{\partial#1}{\partial#2}} 
\newcommand{\pderi}[2]{\partial#1 / \partial#2} 
\newcommand{\tderi}[2]{\mathrm{d}#1 / \mathrm{d}#2} 

\usepackage{makecell}

\makeatletter
\newcommand{\thickhline}{%
    \noalign {\ifnum 0=`}\fi \hrule height 1pt
    \futurelet \reserved@a \@xhline
}
\newcolumntype{"}{@{\hskip\tabcolsep\vrule width 1pt\hskip\tabcolsep}}
\makeatother

\begin{document}

\preprint{{\em Accepted to Physical Review Fluids}}

\title{General method for determining the boundary layer thickness in nonequilibrium flows}

\author{Kevin Patrick Griffin}
 \email{kevinpg@stanford.edu}
\author{Lin Fu}
\author{Parviz Moin}
\affiliation{%
Center for Turbulence Research, Stanford University, Stanford, CA 94305-3024, United States of America
}%

\date{\today}

\begin{abstract}
While the computation of the boundary-layer thickness is straightforward for canonical equilibrium flows, there are no established definitions for general non-equilibrium flows. In this work, a method is developed based on a local reconstruction of the ``inviscid'' velocity profile $U_I[y]$ resulting from the application of the Bernoulli equation in the wall-normal direction. The boundary-layer thickness $\delta_{99}$ is then defined as the location where $U/U_I = 0.99$, which is consistent with its classical definition for the zero-pressure-gradient boundary layers (ZPGBLs). The proposed local-reconstruction method is parameter free and can be deployed for both internal and external flows without resorting to an iterative procedure, numerical integration, or numerical differentiation. The superior performance of the local-reconstruction method over various existing methods is demonstrated by applying the methods to laminar and turbulent boundary layers and two flows over airfoils.
Numerical experiments reveal that the local-reconstruction method is more accurate and more robust than existing methods, and it is applicable for flows over a wide range of Reynolds numbers.
\end{abstract}
%
\maketitle


\section{\label{sec:intro}Introduction}

For both experiments and simulations of viscous flows, it is crucial to be able to define and compute the boundary-layer thickness $\delta$, i.e. the wall-normal distance beyond which viscous effects are negligible. Due to the fact that the outer layer of a turbulent boundary layer and the entire laminar boundary layer are self-similar with respect to $\delta$, it is typically a required input for models of laminar and turbulent flows, e.g.  the Coles' law of the wake \cite{Coles1956} and more recent descriptions of the turbulent velocity profile \cite{Afzal1996,Nickels2004,Cantwell2019}. In addition, defining $\delta$ is necessary to compute integral measures of boundary layer, e.g. the displacement thickness $\delta^*$ and momentum thickness $\theta$ as the length scales associated with the near-wall mass and momentum deficits, and the boundary layer shape factor $H = \delta^*/\theta$, which are useful for the analysis \cite{Spalart1993,Bobke2017,Vinuesa2017a,Asada2018,Volino2020,Uzun2020,SanmiguelVila2020} and modeling \cite{White1969,Bernard2003,Griffin2020} of pressure-gradient flows.

For ZPGBLs, the flow approaches a freestream velocity $U_\infty$ as the wall-normal distance $y \rightarrow \infty$. The boundary-layer thickness $\delta_n$ is defined as the wall-normal distance $y$ at which $n\%$ of $U_\infty$ is attained, i.e.
\begin{equation} \label{eq:delta_n_zpg}
    \frac{U}{U_\infty}\Big|_{y=\delta_{n}} = \frac{n}{100},
\end{equation}
where $n=99$ is typically adopted.

However, the above definitions are not applicable to canonical flows (e.g boundary layer flows) with pressure gradients or flows over complex geometries. Taking the flow over a curved plate for instance, the classical ZPGBL definitions do not correctly delimit the region where viscous effects are important, since the inviscid solution features wall-normal gradients in the mean velocity field.

In more general flows, $U$ does not approach a constant asymptote at the boundary-layer edge.
Once $\delta$ is defined, the boundary-layer edge velocity can be defined as $U_e = U[y=\delta]$, and the displacement and momentum thicknesses can be truncated to only include contributions from the viscous region of the flow.

In this work, various existing methods for defining $\delta$ for general non-equilibrium flows are analyzed and a method based on the application of the Bernoulli equation in the wall-normal direction is proposed.
The remainder of this paper is organized as follows. In section \ref{sec:existing_methods}, the existing methods for computing the boundary-layer thickness are reviewed and their corresponding limitations are analyzed. In section \ref{sec:new_method}, the present method based on stagnation pressure is developed. In section \ref{sec:results}, the performance of the proposed method is compared with existing methods and the shortcomings of existing methods are demonstrated. Concluding discussions and remarks are given in section \ref{sec:conclusion}.

\section{\label{sec:existing_methods} A brief review of existing methods for determining $\delta$}

In this section, several widely used or recently developed methods for computing the boundary-layer thickness are reviewed and the limitations of each method are analyzed. Methods not quantitatively evaluated in this work are relegated to appendices. Specifically, the methods based on composite velocity profiles \cite{Nickels2004,Vinuesa2016} are discussed in Appendix~\ref{app:composite_profiles}, those based on intermittency thresholds \cite{Murlis1982,Jimenez2010,Li2011} are discussed in Appendix~\ref{app:intermittency}, and those based on assumed inviscid profiles are discussed in Appendix~\ref{app:assumed_inviscid_profile}.

\subsection{Mean-vorticity-based methods} \label{sec:mean_vorticity}
Based on the assumption that the freestream flow is irrotational, the mean-vorticity-based methods identify the edge of the rotational viscous boundary layer as the boundary-layer thickness. Recalling Lighthill's \cite{Lighthill1963} `generalized velocity,'
\begin{equation} \label{eq:define_U_tilde}
    \widetilde{U} = \int_0^y -\Omega_z dy',
\end{equation}
where $\Omega_z = \pderi{V}{x}-\pderi{U}{y}$ denotes the mean vorticity component in the spanwise direction,
Spalart and Watmuff \cite{Spalart1993} assert that the integral for $\widetilde{U}_\infty = \widetilde{U}[y \rightarrow \infty]$ converges since the vorticity is assumed to vanish in the inviscid region outside the boundary layer. 
However, since a semi-infinite integration domain is practically not available, an initial guess on the upper limit of integration is required, and it is necessary to iteratively recompute on a larger domain until convergence is established. Nonetheless, this method has been successfully deployed by Tamaki et al. \cite{tamaki2020} for the flow over an airfoil at a near-stall condition.

Two cognate methods for finding both $\delta$ and $U_e$ have been proposed by Coleman et al. \cite{Coleman2018} and Uzun \& Malik \cite{Uzun2020}. 

\subsubsection{The $\widetilde{U}_\infty$ method} 
\begin{figure}
  \centering
  \includegraphics[width=0.5\linewidth]{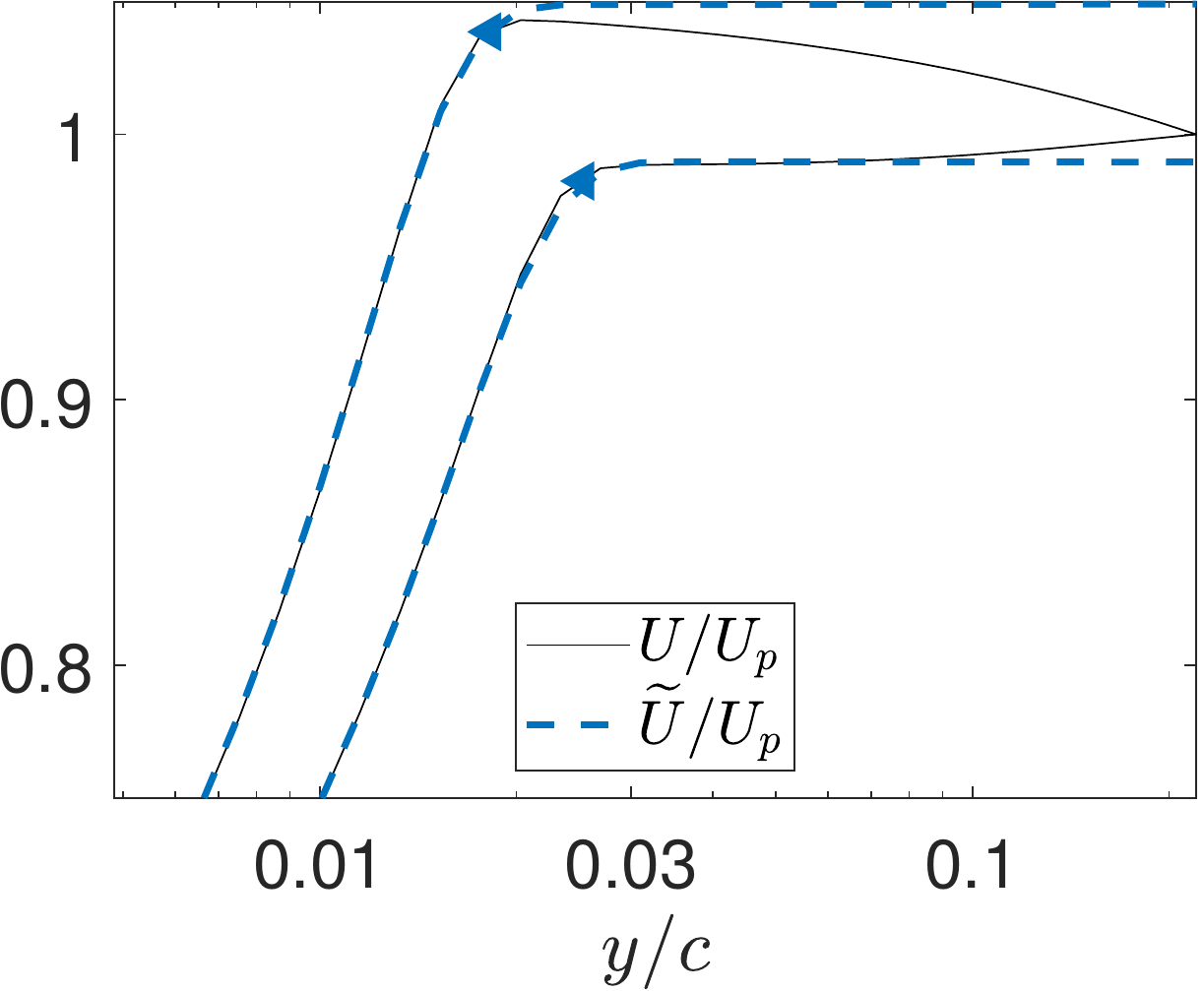}
  \caption{Included are a profile from the suction side of a NACA 4412 airfoil \cite{Vinuesa2018} at the angle-of-attack (AoA) $=5^{\circ}$ and the Reynolds number based on the chord length $Re_c=10^6$ at the streamwise location $x/c = 0.72$ (upper) and a profile from a NACA 0012 airfoil \cite{Tanarro2020} at AoA $=0^{\circ}$ and $Re_c=400,000$ at the streamwise location $x/c =  0.90$ (lower). Distributions of the mean streamwise velocity (black solid lines) from simulations and the generalized velocity (blue dashed lines) computed using Eq.~(\ref{eq:define_U_tilde}) versus the wall-normal coordinate $y$ normalized by the airfoil chord $c$. Blue triangles indicate the locations $y = \delta_{99}$ identified from Eq.~(\ref{eq:delta_n_coleman}). Note that $U_p$ denotes the velocity at the furthest point from the wall in the wall-normal profile.}
  \label{fig:U_tilde_vs_y_demoC}
\end{figure}

The approach of Coleman et al. \cite{Coleman2018} determines $\delta_n$ by finding the wall-normal distance at which 
\begin{equation} \label{eq:delta_n_coleman}
    \frac{\widetilde{U}}{\widetilde{U}_\infty}\Big|_{y=\delta_n} = \frac{n}{100}.
\end{equation}
This approach is referred to as the ``$\widetilde{U}_\infty$ method" hereafter.
In Fig.~\ref{fig:U_tilde_vs_y_demoC}, distributions of the mean streamwise velocities for two cases are plotted as well as those of the corresponding generalized velocities for comparison. $y_p$ denotes the furthest point from the wall in the wall-normal profile; it is an arbitrary length scale. While the mean streamwise velocities continue to vary as $y\rightarrow y_p$, the generalized velocity $\widetilde{U}$ achieves a constant asymptotic value as the flow becomes irrotational. This permits the robust deployment of Eq.~(\ref{eq:delta_n_coleman}) even though $\widetilde{U}_\infty$ is evaluated at a finite wall-normal distance in practice.

\subsubsection{The $-y\Omega_z$ threshold method} 
\begin{figure}
  \centering
  \includegraphics[width=0.5\linewidth]{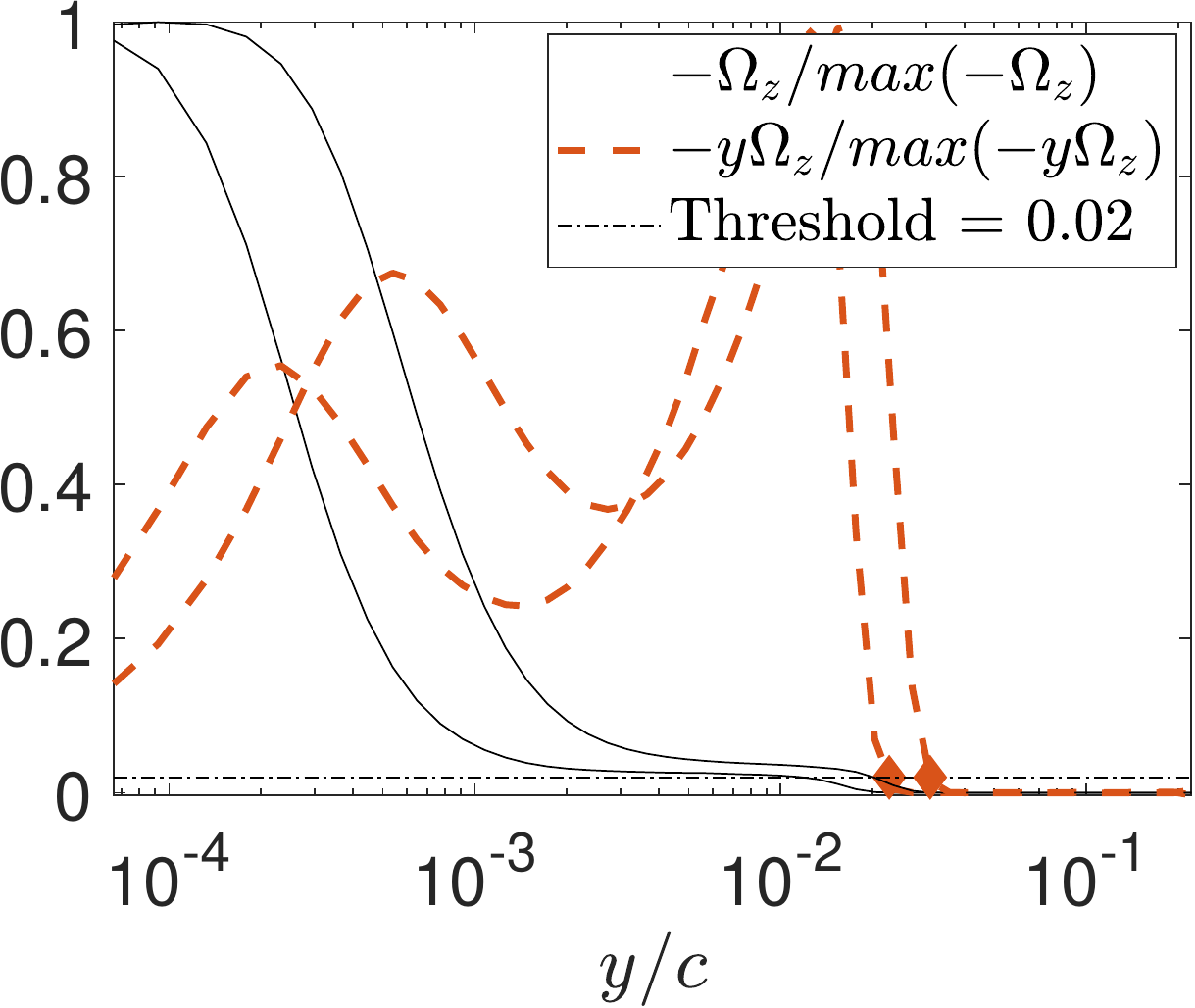}
  \caption{Distributions of the mean z-vorticity $\Omega_z$ (black solid lines) and the boundary layer sensor $-y\Omega_z$ \cite{Uzun2020} (red dashed lines) versus the wall-normal coordinate $y$ normalized by $c$. The red diamonds indicate the locations $y = \delta_{99}$ identified from Eq.~(\ref{eq:delta_n_uzun}). Included are a profile from the suction side of a NACA 4412 airfoil at AoA $=5^{\circ}$ and $Re_c=10^6$ \cite{Vinuesa2018} at the streamwise location $x/c=0.72$ (left curves and symbols) and a profile from a NACA 0012 airfoil at AoA $=0^{\circ}$ and $Re_c=400,000$ \cite{Tanarro2020} at the streamwise location $x/c=0.90$ (right curves and symbols).}
  \label{fig:Omega_z_vs_y_demoC}
\end{figure}

In the approach of Uzun \& Malik \cite{Uzun2020}, a threshold for $-y \Omega_z$ is empirically connected to the desired $\delta_n$, i.e.
\begin{equation} \label{eq:delta_n_uzun}
    \frac{-y \Omega_z}{\max(-y \Omega_z)}\Big|_{y=\delta_n} = C_\Omega,
\end{equation}
where $C_\Omega$ is an empirical parameter that depends on the choice of $n$. For instance, for $\delta_{99}$, $C_\Omega = 0.02$ is suggested \cite{Uzun2020}. This approach is henceforth referred to as the ``$-y\Omega_z$ threshold method." In Fig.~\ref{fig:Omega_z_vs_y_demoC}, distributions of $-\Omega_z$ and $-y\Omega_z$ profiles are plotted versus the wall-normal distance. While $-\Omega_z$ is monotonically decreasing, the profile of $-y\Omega_z$ features multiple peaks and achieves its maximum in the outer portion of the boundary layer. Since the $-y\Omega_z$ profile crosses the threshold $C_\Omega$ typically at two locations, i.e. near the wall (not shown in log scale) and in the vicinity of the boundary-layer edge, the furthest point from the wall that satisfies Eq.~(\ref{eq:delta_n_uzun}) is typically adopted, as suggested by Uzun and Malik \cite{Uzun2020}. The main shortcoming of this method is the sensitivity to the empirical threshold, although an optimal choice may deliver decent results as discussed in section~\ref{sec:results}.

The consistency of mean-vorticity-based methods with the classical ZPGBL definition of $\delta$ depends on the approximation that $\pderi{V}{x} << \pderi{U}{y}$, which is valid in the limit that the boundary-layer thickness $\delta$ is much less than the distance from the flat-plate leading edge, which holds in the limit of large $Re_x = U_e x/ \nu$. Therefore, the deployment of the mean-vorticity methods to flows with low Reynolds numbers is questionable. 

Another disadvantage is that the mean-vorticity-based methods rely on the data of $\pderi{V}{x}$, which is typically not available from existing databases of simulations or experiments although can be reported from future work straightforwardly.

\subsection{Methods based on a mean-shear threshold} \label{sec:mean_shear}
In the mean-shear-based methods, the boundary-layer thickness $\delta$ is defined as the wall-normal distance $y$, where the mean shear $\pderi{U}{y}$ drops below an arbitrary threshold. A special case of this method, henceforth referred to as the ``max method," is when the threshold is set to be zero, such that $y = \delta$ occurs at the maximum of $U[y]$. This method is widely used, and recently is applied to an airfoil at a near-stall condition by Asada and Kawai \cite{Asada2018}. However, this choice of the threshold may never be crossed if the mean streamwise velocity profile is monotonically increasing, e.g., in the flow over a NACA 0012 airfoil in Fig.~\ref{fig:U_tilde_vs_y_demoC}. For general scenarios, a positive threshold is preferred. On the other hand, it is worth noting that the mean shear must be appropriately non-dimensionalized such that the same threshold can be deployed for a wide range of flows.

For laminar flows, an appropriate non-dimensionalization is
\begin{equation} \label{eq:mean_shear_laminar}
    \pder{U/U_\infty}{y/\delta}\Big|_{y=\delta_n} = C_l,
\end{equation}
where $C_l$ is the laminar threshold. Concerning the special case of a Blasius boundary layer, the threshold can be analytically given as the wall-normal distance $y=\delta_n$ satisfying $U/U_\infty = n/100$ (see section~\ref{sec:laminar}).

For turbulent flows, the conventional non-dimensionalization of the mean shear is in wall units, which are defined as the wall shear stress $\tau_w = \tau|_{y=0}$, the wall friction velocity $u_\tau= \sqrt{\tau_w/\rho}$, and the wall viscous unit $\delta_v = \nu/u_\tau$. For this non-dimensionalization, an arbitrary threshold of 0.001 is adopted by Vinuesa et al. \cite{Vinuesa2016}.

However, this non-dimensionalization does not lead to a general threshold constant. Considering the fully turbulent flow over a flat plate with a pressure gradient, the boundary layer can be described by the Coles' wall-wake law \cite{Coles1956} as
\begin{equation} \label{eq:coles}
    U^+ = \frac{1}{\kappa} \ln(y^+) + B + \frac{2\Pi}{\kappa} \sin^2(\frac{\pi y}{2 \delta}),
\end{equation}
where $\kappa$, $B$, and $\Pi$ are empirical parameters. The superscript $+$ refers to quantities non-dimensionalized via $\delta_v$ and $u_\tau$. Examining this correlation near the boundary-layer edge implies that
\begin{equation} \label{eq:wrong_nondim}
    \pder{U^+}{y^+}\big|_{y=\delta_n} \approx \frac{1}{\kappa Re_\tau},
\end{equation}
where the friction Reynolds number $Re_\tau = \delta u_\tau/\nu$. Multiplying Eq.~(\ref{eq:wrong_nondim}) by $Re_\tau$ suggests that the following mixed non-dimensionalization is appropriate for the application of the mean-shear threshold method to turbulent flows, i.e.
\begin{equation}
    \pder{U^+}{y/\delta}\big|_{y=\delta_n} = C_t,
\end{equation}
where $C_t$ is the turbulent mean-shear threshold. This analysis implies that $C_t \approx 1/\kappa$ for velocity profiles that are well described by Eq.~(\ref{eq:coles}).

Theoretically, the mean shear can be decomposed into a viscous part (as discussed above) and an inviscid part resulting from the streamline curvature in the freestream. While both parts decay with wall-normal distance, the viscous part has positive magnitude and decays with a length scale of $\delta$, whereas, the inviscid part can have arbitrary sign and a decay length scale that is typically on the order of the local radius of curvature $R$ of the geometry. The inviscid contribution to the mean shear can be negligible when $\delta << R$, which, however, is not guaranteed for the flow over a thin airfoil at moderate Reynolds number.
In practice, the mean-shear-based methods provide no way to distinguish the viscous and inviscid contributions to the mean shear, such that the above guidelines for setting the threshold based only on the viscous part are not generally applicable. These methods will only succeed when the inviscid contribution is small and the threshold constant is set to be greater than the inviscid part. In other words, for the case with a monotonically increasing velocity profile, there is no guarantee that a given threshold will be crossed.

This observation motivates the present method proposed in section~\ref{sec:new_method}, which relies on the local reconstruction of the inviscid solution. Moreover, the local-reconstruction method works directly with the velocity profile instead of the mean shear due to the fact that the mean shear is comparatively more sensitive to noise in the velocity profile. 

Technically, there exist two additional drawbacks for the mean-shear-based methods. (i) Since both the laminar and the mixed non-dimensionalizations depend on $\delta$, $\delta$ must be computed iteratively, unless the threshold constant is selected to be zero. 
It is worth noting that for a general mean velocity profile, there is no guarantee that these iterative methods will converge. (ii) Although, in principle, a reliable choice for $C_l$ and $C_t$ may exist based on the dimensional analysis above for the fully-turbulent and fully-laminar boundary layers, respectively, the extension for transitional flows is challenging since neither non-dimensionalization is valid in the transitional regime.

\subsection{The diagnostic-plot method} \label{sec:diagnostic_plot}
Vinuesa et al. \cite{Vinuesa2016} propose a novel approach for determining the boundary-layer thickness, which is based on the diagnostic plot \cite{Alfredsson2011} of the streamwise turbulence intensity $\sqrt{\overline{u'^2}}/U_e$ versus the mean streamwise velocity $U/U_e$. Alfredsson et al. \cite{Alfredsson2011} claim that the diagnostic plot becomes nearly universal as $y\rightarrow \delta$. Moreover, Dr{\'o}zdz et al. \cite{Drozdz2015} assert that the diagnostic plot can be generalized to non-equilibrium flows by considering $\sqrt{\overline{u'^2}}/(U_e \sqrt{H})$ versus $U/U_e$. Indeed, the collapse of various velocity profiles in Fig.~4 of \cite{Vinuesa2016} is impressive even though the theoretical basis for the collapse has not yet been established. Based on the observed asymptotic behavior of the diagnostic plot, Vinuesa et al. \cite{Vinuesa2016} further define the boundary-layer edge $\delta_{99}$ as the wall-normal distance at which $\sqrt{\overline{u'^2}}/(U_e \sqrt{H})=0.02$. However, since the criterion for determining $\delta$ depends on the boundary-layer edge velocity $U_e$, an iterative procedure and an appropriate initial guess for $\delta$ are required.
 
One advantage of this method is that it only requires $\sqrt{\overline{u'^2}}$ and $U$ statistics as inputs, which are commonly available in existing databases of experiments and direct numerical simulations (DNS). However, for large-eddy simulations (LES), the turbulence intensity $\sqrt{\overline{u'^2}}$ (unlike the mean velocity $U$) is typically under-resolved with coarse meshes and features strong resolution dependence. Consequently, the boundary-layer thickness estimate from the diagnostic-plot method depends on the LES mesh resolution. On the other hand, in a Reynolds-averaged Navier-Stokes (RANS) simulation, the turbulence intensity $\sqrt{\overline{u'^2}}$ is modeled with built-in modeling errors, rendering the performance of the diagnostic-plot method dependent on the RANS-model.

The main shortcoming of the diagnostic-plot method is that the threshold constant of 0.02 is empirical. For an external flow with freestream turbulence or an internal flow, to find an optimal threshold constant may be rather challenging if it does exist. Similarly, The threshold needs to be properly re-calibrated for determining $\delta_{n}$ with $n\ne99$. Moreover, this approach is unsuitable for laminar flows since the turbulence intensity is zero across the entire boundary layer. In other words, the deployment of diagnostic-plot method should be restricted to fully turbulent boundary layers with a quiet freestream.

\section{\label{sec:new_method} Local-reconstruction method for identifying the boundary-layer thickness}

In this section, we propose a method to identify the boundary-layer thickness, which (i) does not require numerical integration, numerical differentiation, or empirical thresholds, (ii) has a guaranteed solution without resorting to an iterative procedure, (iii) is applicable at all Reynolds numbers for internal and external flows with or without freestream turbulence, and (iv) only relies on the mean velocity profiles $U[y]$ and $V[y]$, and the mean pressure profile $P[y]$ as inputs. 

The proposed new definition for the boundary-layer thickness is given by
\begin{equation} \label{eq:delta_n_new}
    \frac{U}{U_I}\Big|_{y = \delta_{n}} = \frac{n}{100},
\end{equation}
where $U_I[y]$ denotes a local reconstruction of the inviscid mean streamwise velocity profile. 
This method is a generalization of the ZPG definition for $\delta$ given by Eq.~(\ref{eq:delta_n_zpg}). The two definitions are equivalent for a ZPGBL, where the inviscid solution is simply a constant, i.e., $U_I[y] = U_\infty$.

Formally, the inviscid velocity profile can be obtained by solving the Euler equations and imposing no-penetration boundaries contoured to the local displacement thickness computed in a viscous simulation. However, since the displacement thickness depends on the boundary-layer thickness, this approach is impractical. On the other hand, in this work, a simpler and more computationally efficient local reconstruction of the inviscid solution $U_I[y]$ is proposed. The approach begins by defining the stagnation pressure for an incompressible steady flow as
\begin{equation} \label{eq:bernoulli}
   P_o = P + \frac{1}{2} \rho U_m^2,
\end{equation}
where $P$ is the static pressure and the velocity magnitude squared $U_m^2 = U_I^2 + V^2$ is computed from the reconstructed mean streamwise velocity $U_I$ and the mean wall-normal velocity $V$ in the 2D local reference frame.

While the no-penetration boundary condition, i.e., $V[y=0]=0$, applies for both inviscid and viscous flows, the no-slip boundary condition, i.e., $U[y=0] = 0$, applies only to viscous flows. As a result, the reconstructed inviscid mean streamwise velocity profile $U_I[y]$ deviates from $U[y]$ near the wall, where the viscous effects dominate. Meanwhile, the wall-normal velocity profile $V[y]$ and the mean pressure profile $P[y]$ feature a much weaker dependence on viscous effects. For a flat-plate pressure-gradient boundary layer, $\pderi{P}{y} \approx 0$ and $V\approx 0$ in both the viscous and the inviscid cases.

Considering a hypothetical irrotational, inviscid flow that has $V[y]$ and $P[y]$ profiles equivalent to those of the corresponding viscous flow, the Bernoulli equation may approximately apply globally rather than just along streamlines. Therefore, the corresponding streamwise velocity $U_I[y]$ is obtained from Bernoulli's equation following 
\begin{equation} \label{eq:U_I_bernoulli}
    U_I = \pm\sqrt{\frac{2}{\rho}\left(P_{o,ref} - P[y]\right) - V^2[y]},
\end{equation}
where $P_{o,ref}$ denotes the total pressure at a reference location specified below. The sign of $U_I$ should be chosen to match that of $U[y=\delta]$. (In practice, it is recommended for attached flows that the boundary-layer thickness be computed from the squares of Eq.~(\ref{eq:U_I_bernoulli}) and Eq.~(\ref{eq:delta_n_new}) to dispense with the square root operator).

\begin{figure}
\begin{subfigure}{.5\textwidth}
  \centering
  \includegraphics[width=1\linewidth]{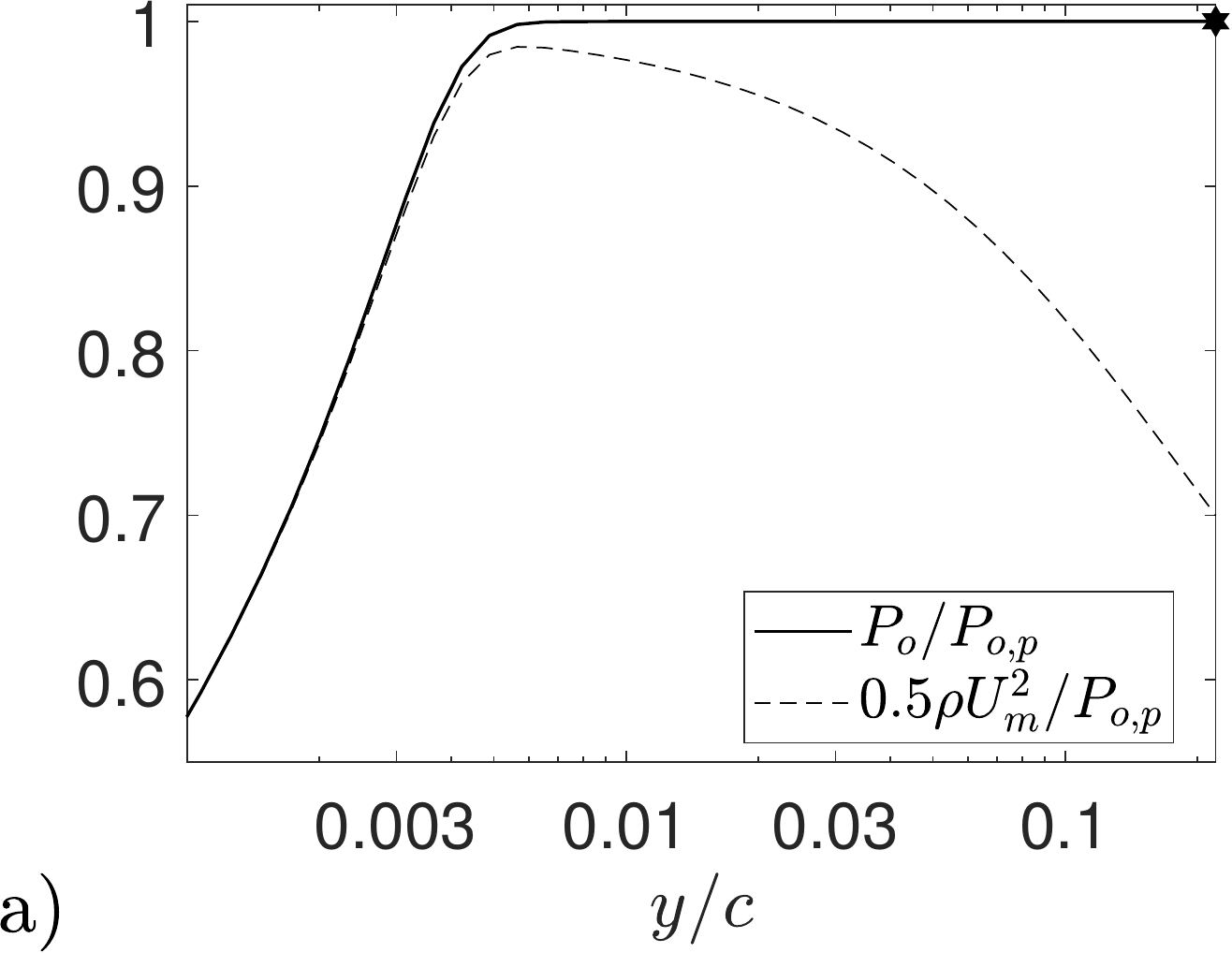}
  \captionlistentry{}
  \label{fig:Po_vs_y_demoA}
\end{subfigure}%
\begin{subfigure}{.5\textwidth}
  \centering
  \includegraphics[width=1\linewidth]{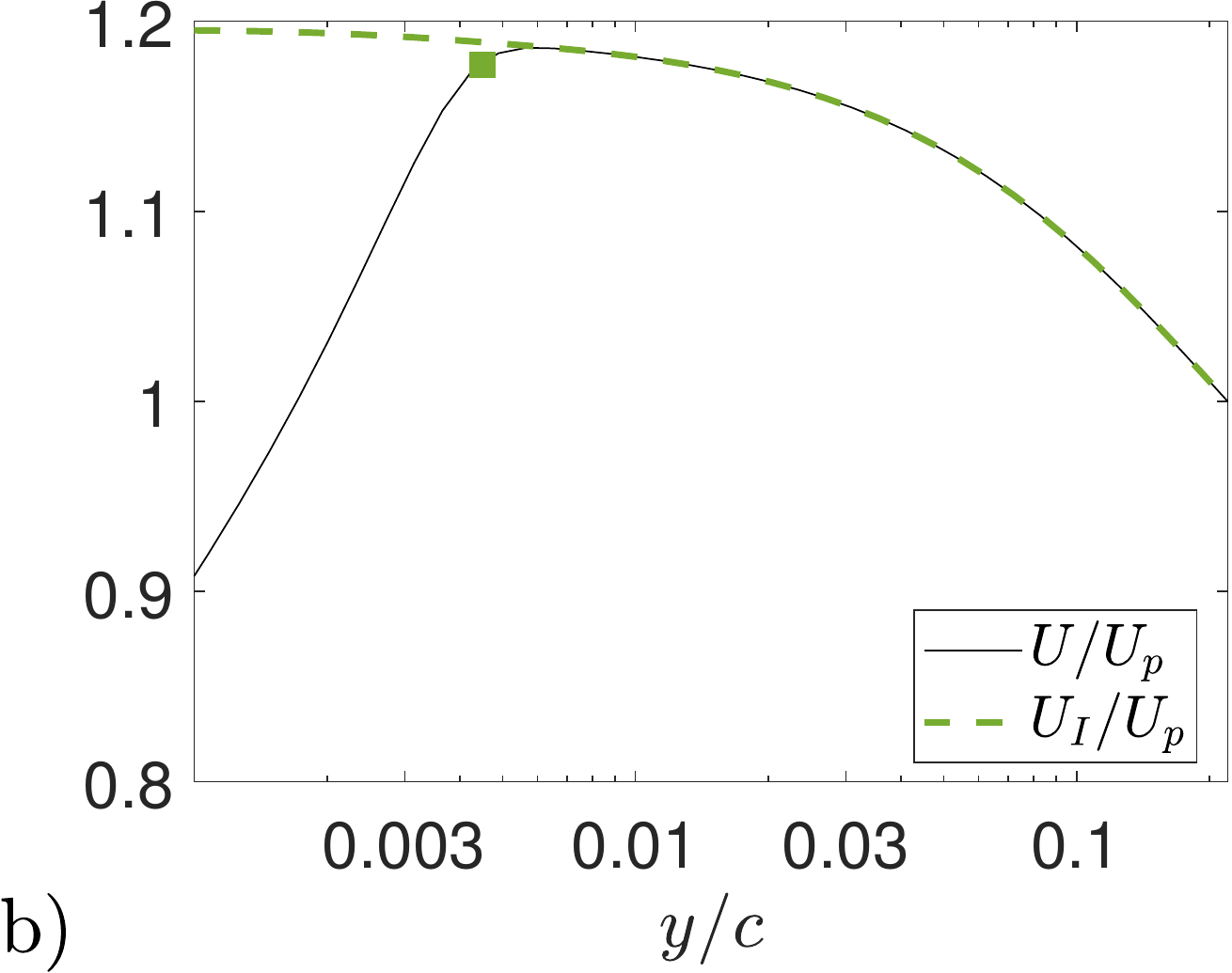}
  \captionlistentry{}
  \label{fig:UI_vs_y_demoA}
\end{subfigure}%
\caption{Distributions of the stagnation pressure $P_o$ and the dynamic pressure (a), and the velocity profile $U$ and the locally reconstructed inviscid velocity profile $U_I$ (b) versus the normalized wall-normal coordinate $y/c$. The maximum stagnation pressure (a) and the boundary-layer edge (b) are indicated with the black hexagram and the green square, respectively. The data is from the suction side of a NACA 4412 airfoil at AoA $=5^{\circ}$ and $Re_c=10^6$ \cite{Vinuesa2018} at the streamwise station $x/c=0.20$.}
\label{fig:demoA_Po_and_UI}
\end{figure}

The rationale is that the viscous effects are approximately confined to the boundary layer, i.e. the flow is nearly inviscid and irrotational outside of the boundary layer with $y>\delta$. This implies that $P_{o,ref}$ is nearly a constant across the domain with $y>\delta$. Taking the flow over an airfoil for instance, as shown in Fig.~\ref{fig:Po_vs_y_demoA}, the dynamic pressure varies continuously whereas the stagnation pressure preserves a constant asymptote outside of the boundary layer. Meanwhile, one unique choice of $P_{o,ref}$ is given by setting $P_{o,ref} = \max (P_o)$, which is achieved at a wall-normal distance $y>\delta$. Since $P_o$ is a measure of the flow's capacity to do work, it typically decreases as a wall is approached. This choice of $P_{o,ref}$ guarantees that the model has the correct behavior in channel and pipe flows, as will be discussed in section~\ref{sec:internal_flows}.

The locally reconstructed inviscid solution $U_I[y]$ given by Eq.~(\ref{eq:U_I_bernoulli}) is plotted for the flow over a NACA 4412 airfoil \cite{Vinuesa2018} in Fig.~\ref{fig:UI_vs_y_demoA}. As expected, the inviscid solution $U_I[y]$ has an excellent agreement with the viscous solution $U[y]$ outside the boundary layer. The boundary-layer thickness $\delta_{n}$ is then estimated to be the location where $U_I$ departs from $U$ by $100-n$ percent. In this work, the boundary-layer edge velocity is taken to be $U_e = U[y=\delta_{n}]$. The choices of $U_e = U_I[y=\delta_{n}]$ or $U_e = U_I[y=0]$ are also reasonable \cite{Kim1991,Uzun2020}.

\begin{figure}
\begin{subfigure}{.5\textwidth}
  \centering
  \includegraphics[width=1\linewidth]{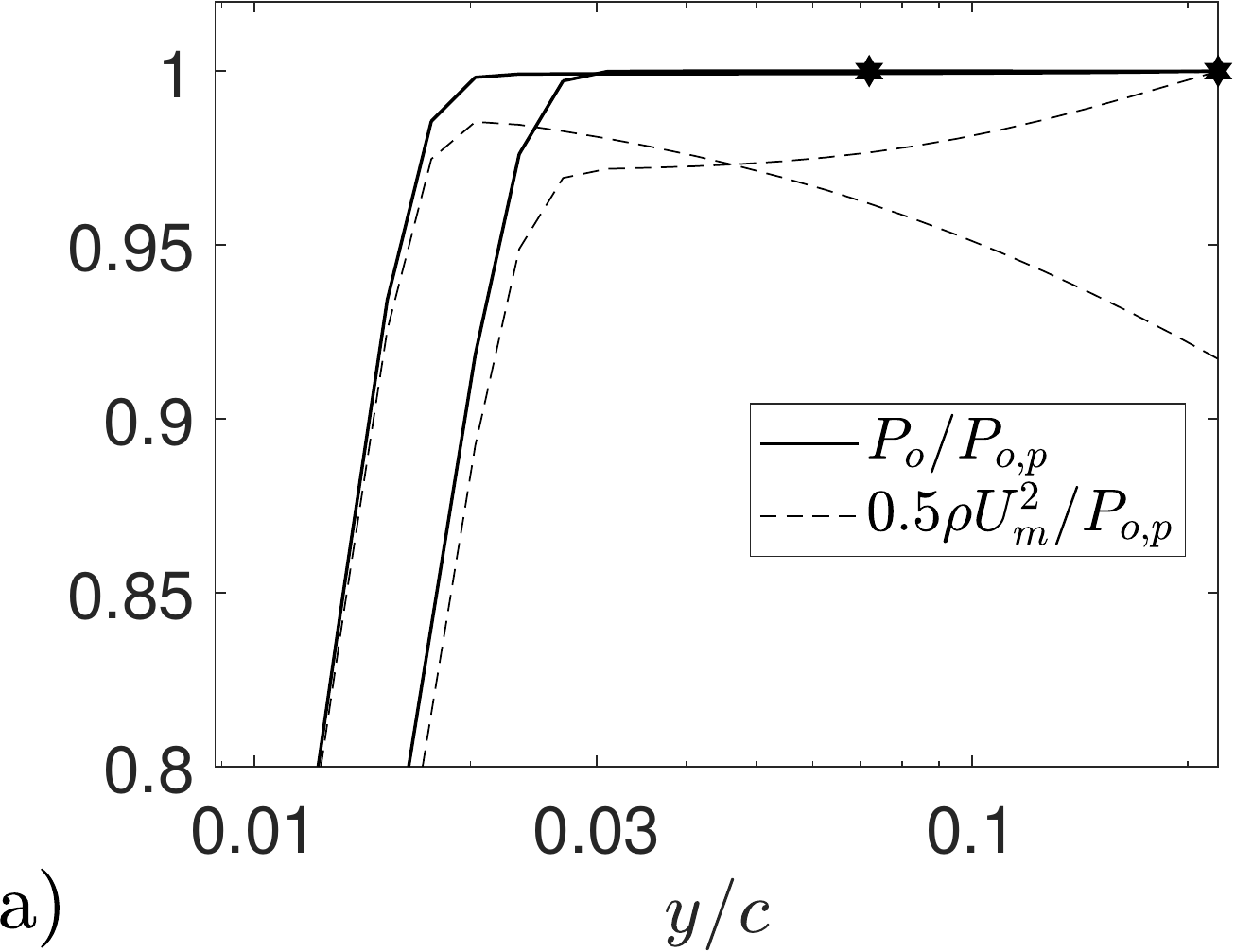}
  \captionlistentry{}
  \label{fig:Po_vs_y_demoC}
\end{subfigure}%
\begin{subfigure}{.5\textwidth}
  \centering
  \includegraphics[width=1\linewidth]{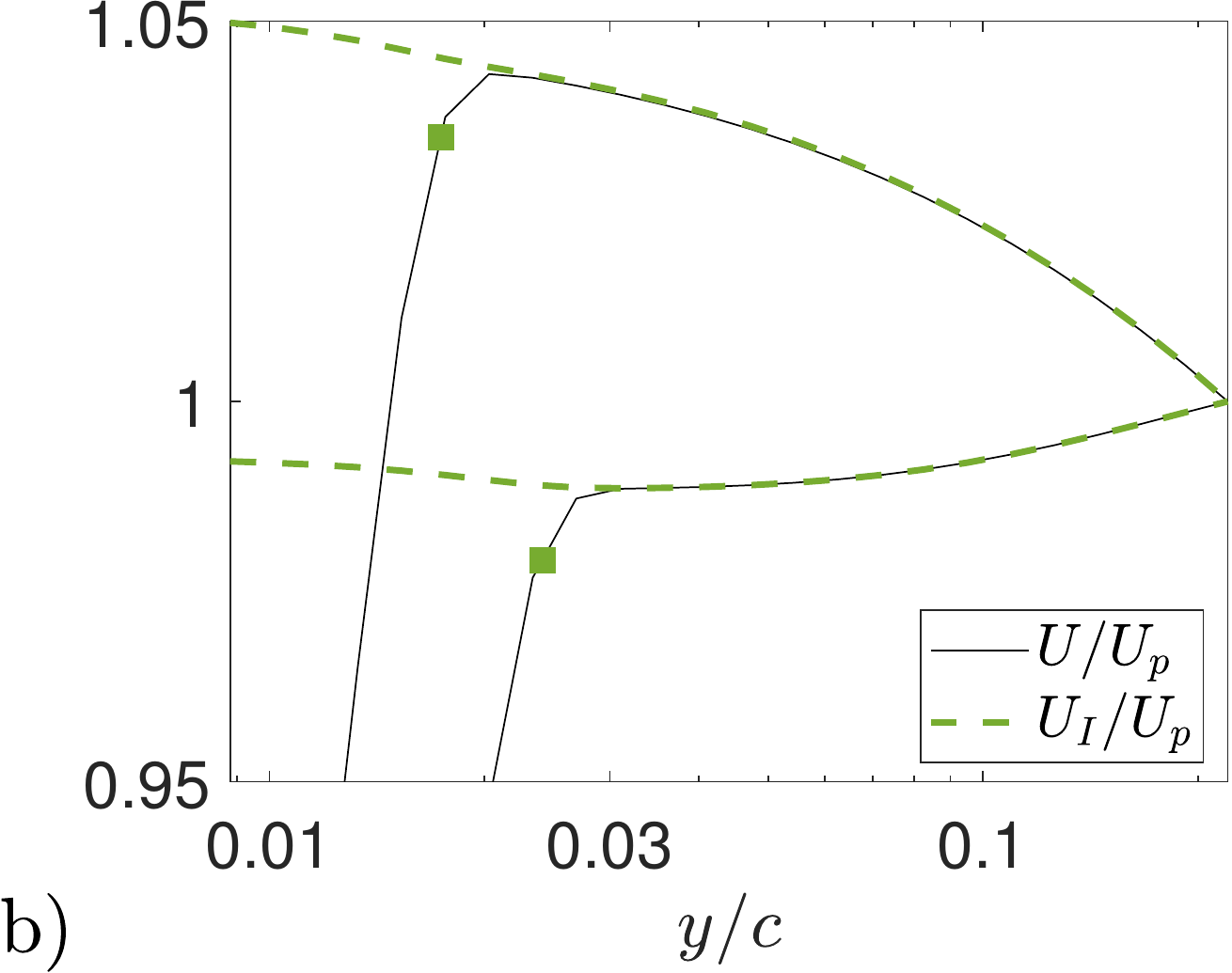}
  \captionlistentry{}
  \label{fig:UI_vs_y_demoC}
\end{subfigure}%
\caption{Distributions of the stagnation pressure $P_o$ and the dynamic pressure (a), and the velocity profile $U$ and the locally reconstructed inviscid velocity profile $U_I$ (b) versus the normalized wall-normal coordinate $y/c$. This is the same as Fig.~\ref{fig:demoA_Po_and_UI} except that one profile is from the suction side of a NACA 4412 airfoil at AoA $=5^{\circ}$ and $Re_c=10^6$ \cite{Vinuesa2018} at the streamwise station $x/c = 0.72$ (left curves and symbols) and the other profile is from a NACA 0012 airfoil at AoA $=0^{\circ}$ and $Re_c=400,000$ \cite{Tanarro2020} at the streamwise station $x/c =  0.90$ (right curves and symbols). The maximum stagnation pressure (a) and the boundary-layer edge (b) are indicated with the black hexagram and the green square, respectively.}
\end{figure}

As shown in Fig.~\ref{fig:Po_vs_y_demoC} and Fig.~\ref{fig:UI_vs_y_demoC}, the local-reconstruction method is successfully deployed to two additional challenging cases, one of which the velocity profile is monotonically increasing. It is observed that the total pressure achieves a maximum near the boundary-layer edge and the locally reconstructed inviscid solution agrees well with the viscous solution outside the boundary layer. The estimates of the boundary-layer thickness are qualitatively consistent with the results in Fig.~\ref{fig:UI_vs_y_demoA}. More quantitative assessments of the performance of the local-reconstruction method will be presented in section~\ref{sec:results}.

Note that, in experiments and coarse simulations alike, small spurious oscillations in the velocity profile can occur. For this reason, the search for $U/U_I = n/100$ should begin at the wall and proceed in the wall-normal direction. As with other methods for computing $\delta_n$, the value of $n$ must be selected reasonably. $n=99$ is typically used but this value can be further reduced to ensure robustness in the presence of noisy data.
The present method can be readily generalized for compressible flows and the reader is referred to Appendix~\ref{app:compressible} for the details.

A related method computes $\delta_n$ based on a threshold for the total pressure \cite{Bradshaw1977}. However, the present method is preferable since it works directly with velocity and therefore can recover the classical definition of $\delta_n$ in a ZPG boundary layer and is not sensitive to a rescaling of the magnitude of pressure.
\section{\label{sec:results} Performance assessments of the proposed and the existing methods}

In this section, the methods discussed in sections~\ref{sec:existing_methods} and~\ref{sec:new_method} will be assessed by deploying to the analytical Blasius boundary layer and the well-resolved numerical simulation database in Table~\ref{tab:database}. 

\begin{table}[]
\begin{tabular}{lcccccc}
\hline
\multicolumn{1}{c}{Type of flow} & $Re_\tau$ & $1000\alpha$ & $\beta$ & $H$ & Sources \\ \hline
Turbulent ZPGBLs & 	[446,2479]	 & [ 0.00, 0.00] & [ 0.00, 0.00] & [1.36, 1.46] & \cite{Sillero2013,Spalart1988,Eitel-Amor2014} \\
NACA 4412 suction side AoA $=5^{\circ}$ & 	[290, 679]	 & [ 1.45, 61.7] & [ 0.072, 13.7] & [1.49, 1.92] & \cite{Vinuesa2018} \\
NACA 0012 AoA $=0^{\circ}$ & 	[131, 371]	 & [ 2.23, 18.8] & [ 0.061, 1.71] & [1.55, 1.69] & \cite{Tanarro2020} \\
NACA 4412 pressure side AoA $=5^{\circ}$ & 	[198, 583]	 & [-1.32,-0.11] & [-0.004,-0.14] & [1.41, 1.56] & \cite{Vinuesa2018} \\
Turbulent APGBL & 	[202, 920]	 & [-2.34, 31.3] & [-0.570, 4.43] & [1.57, 1.91] & \cite{Bobke2017} \\
\hline
\end{tabular}
\caption{Well-resolved simulation database from various flows for model evaluation. Included are ZPGBLs, two flows over airfoils with specified AoA, and an adverse pressure gradient boundary layer (APGBL) with six different pressure gradient conditions. The ranges of friction Reynolds number $Re_\tau$, the inner pressure-gradient parameter $\alpha =  (\delta_v /\tau_w) \tderi{P}{x}$, the outer pressure-gradient parameter $\beta = (\delta^* / \tau_w) \tderi{P}{x}$, and the boundary layer shape factor $H$ are also provided.}
\label{tab:database}
\end{table}

\subsection{\label{sec:laminar} Analysis of a laminar boundary layer}

Since there is no well-established definition of the boundary-layer thickness in complex flows, it is instructive to first assess the methods by considering the laminar Blasius boundary layer, for which there is an established definition of the boundary-layer thickness given by Eq.~(\ref{eq:delta_n_zpg}).

Recalling the Blasius similarity variable
\begin{equation}
    \eta = y \sqrt{\frac{U_\infty}{2 \nu x}},
\end{equation}
where $x$ denotes the distance from the flat-plate leading edge, and defining the non-dimensional function $f[\eta]$ such that
\begin{equation} \label{eq:blasius_U}
    \frac{U}{U_\infty} = f'[\eta],
\end{equation}
the laminar boundary layer equations reduce to the Blasius ordinary differential equation (ODE), i.e. $f''' + f f''= 0$.
To close the ODE, the boundary conditions are $f[0]=f'[0]=0$ and $f'[\eta \rightarrow \infty] \rightarrow 1$. On the other hand, the continuity equation implies that the wall-normal velocity profile is given by,
\begin{equation} \label{eq:blasius_V}
    \frac{V}{U_\infty} = \frac{(\eta f' - f)}{\sqrt{2 Re_x}},
\end{equation}
where $Re_x = U_\infty x / \nu$.

As discussed above, only four of the introduced methods are suitable for laminar problems, i.e. the $\widetilde{U}_\infty$ method, the $-y\Omega_z$ threshold method, the mean-shear threshold method, and the local-reconstruction method.

\subsubsection{The $\widetilde{U}_\infty$ method applied to a laminar boundary layer}
For the $\widetilde{U}_\infty$ method, the definition of the generalized velocity in Eq.~(\ref{eq:define_U_tilde}) can be non-dimensionalized as
\begin{equation} \label{eq:mean_vort_blasius}
    \frac{\widetilde{U}}{U_\infty} = \int_0^y \left( \pder{(U/U_\infty)}{(y'/\delta)} - \pder{(V/U_\infty)}{(x/\delta)} \right) d(y'/\delta).
\end{equation}
Substituting the Blasius velocity profiles from Eq.~(\ref{eq:blasius_U}) and Eq.~(\ref{eq:blasius_V}) into Eq.~(\ref{eq:mean_vort_blasius}) leads to the following relation
\begin{equation} \label{eq:blasius_U_tilde}
    \frac{\widetilde{U}}{U_\infty} = \int_0^\eta \left( f'' + \frac{\eta f' - f}{Re_x} \right) d\eta.
\end{equation}
The $\widetilde{U}_\infty$ method (see Eq.~(\ref{eq:delta_n_coleman})) assumes that the integral in Eq.~(\ref{eq:blasius_U_tilde}) converges. The limit of the integrand as $\eta\rightarrow \infty$ can be evaluated as
\begin{equation}
    \lim_{\eta \rightarrow \infty} (f'' + \frac{\eta f' - f}{Re_x}) = \frac{1.217}{Re_x},
\end{equation}
based on the Blasius boundary conditions, i.e. $f'[\eta \rightarrow \infty] \rightarrow 1$, $f''[\eta \rightarrow \infty] \rightarrow 0$, and $f[\eta \rightarrow \infty] \rightarrow \eta + C_\infty$, where $C_\infty \approx -1.217$ is a universal constant \cite{White2006}. The integrand in Eq.~(\ref{eq:blasius_U_tilde}) does not vanish in the limit $\eta\rightarrow \infty$ and approaches the $Re_x$ dependent value of ${1.217}/{Re_x}$. Therefore, this method is incompatible with the laminar boundary layer equations since the solution of the Blasius equations has finite z-vorticity $\Omega_z = -1.217/Re_x$ in the freestream whereas the $\widetilde{U}_\infty$ method assumes the z-vorticity vanishes in the freestream.

For illustrative purposes, the $\widetilde{U}_\infty$ method can be applied to the laminar boundary layer if the integral for $\widetilde{U}_\infty$ is truncated. Using the exact Blasius solution, the integral is artificially truncated at $\delta_{99}$ in this application.

\subsubsection{The $-y\Omega_z$ threshold method applied to a laminar boundary layer}
The $-y\Omega_z$ threshold method given in Eq.~(\ref{eq:delta_n_uzun}) can be written in terms of the Blasius solution as
\begin{equation}
    \frac{\eta f'' + \eta^2 f'/Re_x - \eta f/Re_x}{\max(\eta f'' + \eta^2 f'/Re_x - \eta f/Re_x)}\Big|_{\eta=\eta_{\delta_n}}= 0.02.
\end{equation}
Although there are typically at least two solutions to Eq.~(\ref{eq:delta_n_uzun}), below $Re_x \approx 610$, there is only one solution near the wall and no solution near the boundary-layer edge due to the overly large threshold constant. In other words, below this critical Reynolds number, the method breaks down because the vorticity decays to a finite value instead of vanishing completely as expected, with a similar reason detailed in the previous sub-section.

\subsubsection{The mean-shear method applied to a laminar boundary layer}
When the mean-shear method is applied to the Blasius boundary layer, the laminar non-dimensionalization in outer units, given in Eq.~(\ref{eq:mean_shear_laminar}), is applicable. Expressing this equation in terms of the Blasius solution leads to
\begin{equation}
    \pder{U/U_\infty}{y/\delta}\Big|_{y=\delta_n} = 
    \eta_{\delta_n} f_{\delta_n}''= C_l,
\end{equation}
where $\eta_{\delta_n}$ and $f_{\delta_n}''$ are the values of $\eta$ and $f''$ evaluated at the boundary-layer edge. From the classical definition of the boundary-layer thickness in Eq.~(\ref{eq:delta_n_zpg}) and the definition of $f'$ in Eq.~(\ref{eq:blasius_U}), $y = \delta_n$ occurs when $f' = n/100$. By consulting the tabulated Blasius solution (see, e.g. \cite{White2006}), $\eta_{\delta_{99}} \approx 3.5$, $f_{\delta_{99}}''\approx 0.02$, implying that $C_l \approx 0.07$ corresponds to $n=99$. While the optimal choice of the parameter $C_l$ can be derived for Blasius boundary layers, it may vary in more general laminar flows.

\subsubsection{The local-reconstruction method applied to a laminar boundary layer}
Since the boundary layer approximation implies that $\pderi{P}{y} \approx 0$, the maximum stagnation pressure $P_{0, max} = P_{0,\infty}$. As such, the locally reconstructed inviscid solution from Eq.~(\ref{eq:U_I_bernoulli}) can be expressed as
\begin{equation}
    U_I[y] = \pm \sqrt{\left(U_\infty^2 + V_\infty^2 \right) - V^2[y]}.
\end{equation}
The positive branch is adopted since the Blasius solution $U[y]$ is always positive. Substituting Eq.~(\ref{eq:blasius_V}) leads to
\begin{equation}
    U_I[y] = U_\infty \sqrt{1+ \left(\frac{(\eta_\infty f'_\infty - f_\infty)}{\sqrt{2 Re_x}}\right)^2 - \left(\frac{(\eta f' - f)}{\sqrt{2 Re_x}}\right)^2}.
\end{equation}
After evaluating $\lim_{\eta \rightarrow \infty} (\eta f' - f) $ as before, the expression
\begin{equation} \label{eq:U_I_new_method_blasius}
    U_{I} = U_\infty \sqrt{1 + \frac{1.217^2 - (\eta f' - f)^2}{2 Re_x} }
\end{equation}
holds. A Laurent series expansion of Eq.~(\ref{eq:U_I_new_method_blasius}) indicates that as $Re_x \rightarrow \infty$, $(U_I - U_\infty)/U_\infty \sim 1/Re_x$. Consequently, considering the new definition for $\delta_n$ (given in Eq.~(\ref{eq:delta_n_new})) based on $U_I$ and the classical definition (given in Eq.~(\ref{eq:delta_n_zpg})) based on $U_\infty$, their difference decays like $1/Re_x$.

\begin{figure}
  \centering
  \includegraphics[width=0.5\linewidth]{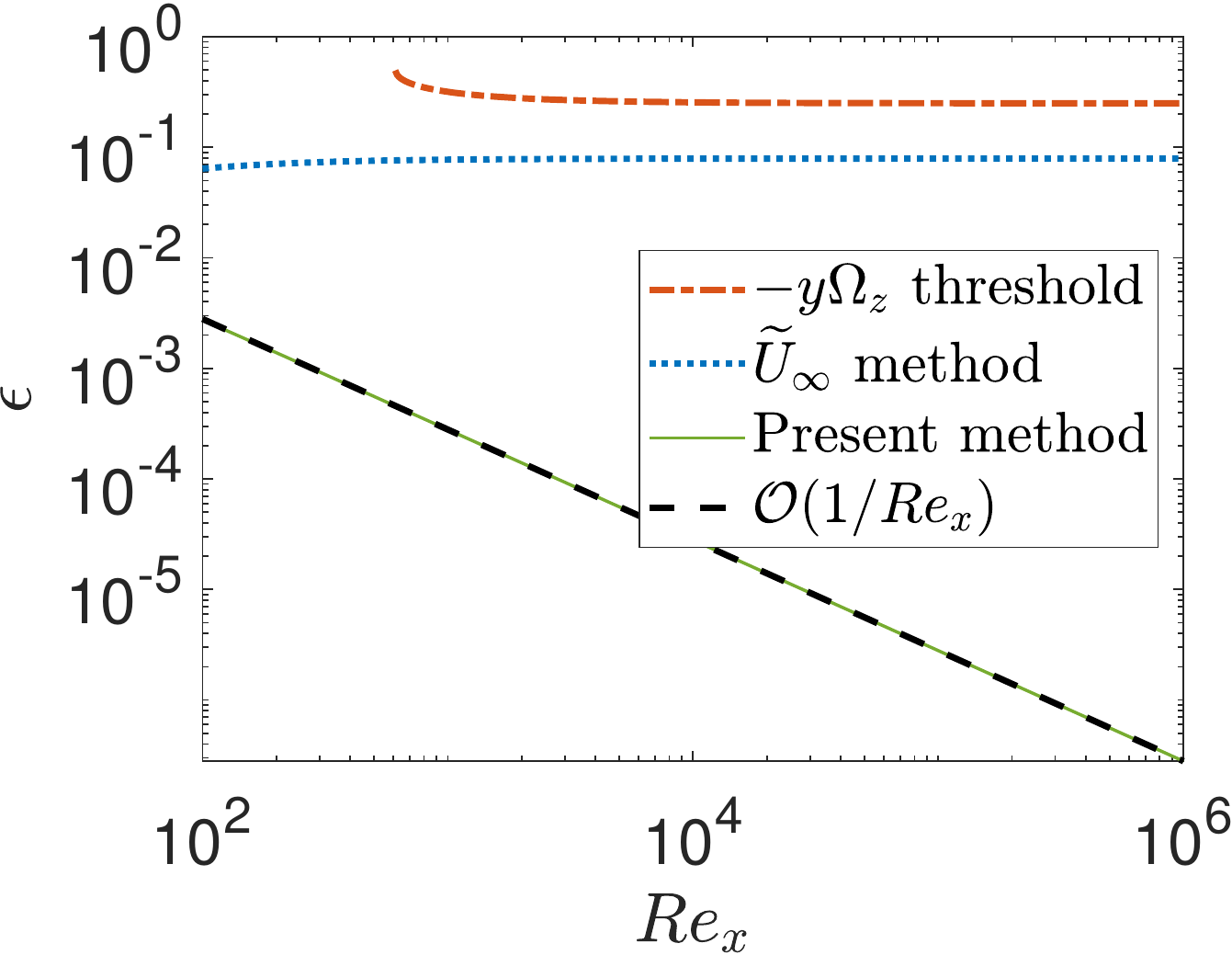}
  \caption{Distributions of the relative error between the computed and the classical definitions of the boundary-layer thickness versus the Reynolds number $Re_x$ for the Blasius boundary layer. Included are the results from the $\widetilde{U}_\infty$ method (blue dotted line), the $-y\Omega_z$ threshold method (red dash-dotted line), and the local-reconstruction method (green solid line) for computing $\delta$ and a $1/Re_x$ reference (black dashed line). Note that the integral in the $\widetilde{U}_\infty$ method is artificially truncated at the analytical boundary-layer edge so that a converged solution is possible for this flow.}
   \label{fig:err_vs_Rex_blasius}
\end{figure}

In Fig.~\ref{fig:err_vs_Rex_blasius}, the relative errors between the computed and the classical definitions of the boundary-layer thickness for the Blasius boundary layer are plotted versus the Reynolds number $Re_x$ between $100$ and $10^6$. It is observed that the error from the $\widetilde{U}_\infty$ method remains almost constant for the entire Reynolds number range because its integral must be artificially truncated. The error from the $-y\Omega_z$ threshold method is relatively large because the empirical threshold constant suggested for turbulent flows \cite{Uzun2020} is inappropriate for laminar boundary layers. In fact, as mentioned above, the boundary-layer thickness is not identified in the regime with $Re_x < 610$. Note that the mean-shear method is exact in the laminar boundary layer and its error therefore is not shown in Fig.~\ref{fig:err_vs_Rex_blasius}. The error from the local-reconstruction method is significantly lower than those from both of the mean-vorticity methods and decays as $1/Re_x$ as expected. 

As a result, both the local-reconstruction method and the mean-shear threshold method are appropriate for deployment in laminar boundary layers. However, the local-reconstruction method is preferred over the mean-shear method since the local-reconstruction method does not require calibration using the exact solution and thus is more generally applicable.

\subsection{\label{sec:turb_zpgbl} Analysis of the zero-pressure-gradient turbulent boundary layer}
In this section, the above discussed methods for computing the boundary-layer thickness are applied to the turbulent ZPGBL, which, like the Blasius boundary layer, has an agreed-upon definition of $\delta$ given by Eq.~(\ref{eq:delta_n_zpg}).

\begin{figure}
  \centering
  \includegraphics[width=0.5\linewidth]{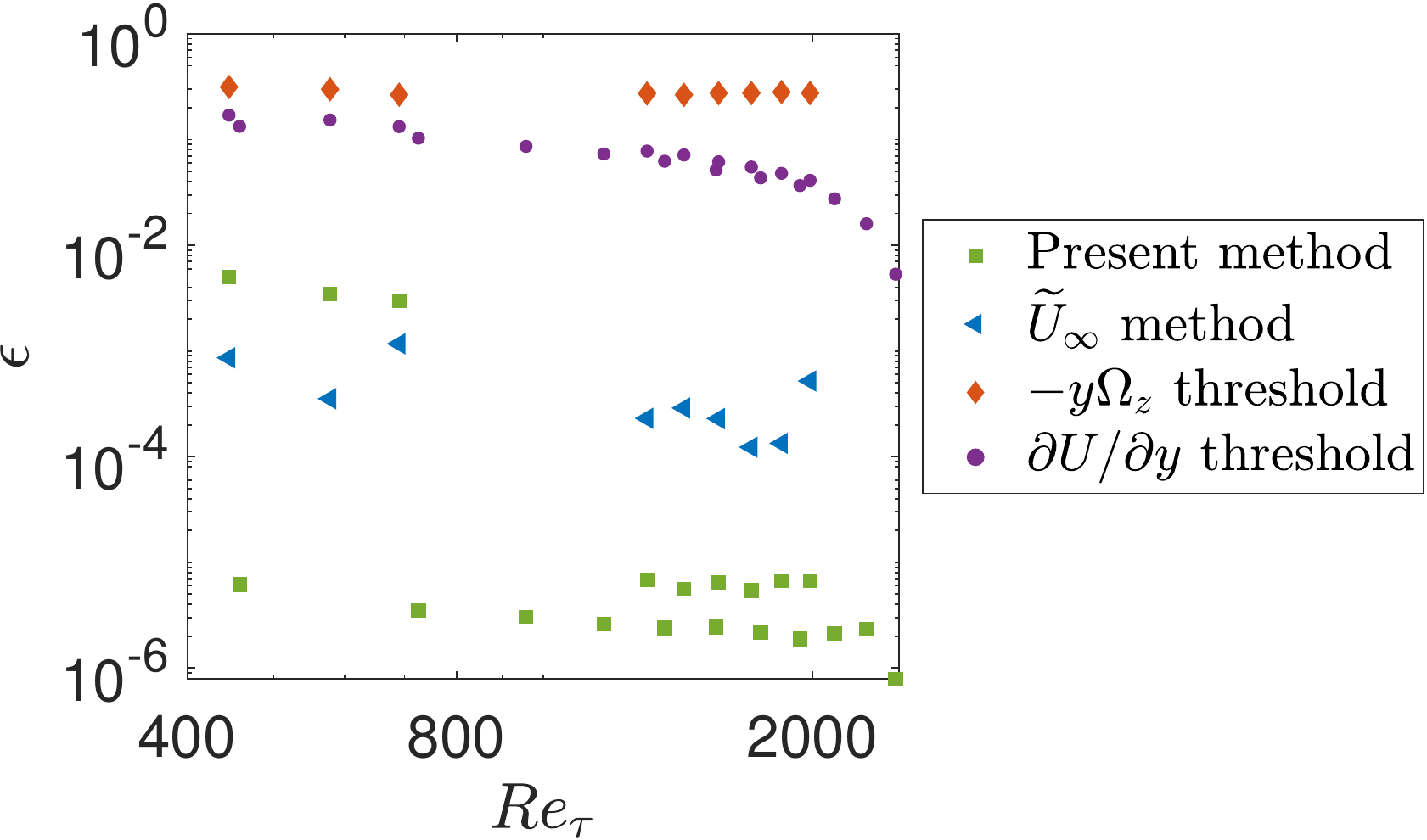}
  \caption{Distributions of the relative error between the computed and the classical definitions of the boundary-layer thickness versus the friction Reynolds number $Re_\tau$. Included are the results from the local-reconstruction method (green squares), the $\widetilde{U}_\infty$ method (blue triangles), the $-y\Omega_z$ threshold method (red diamonds), and the mean-shear threshold method (purple circles). The cases plotted are ZPGBLs of \cite{Jimenez2010,Sillero2013,Eitel-Amor2014}. The mean-vorticity-based methods can not be deployed with the data of \cite{Eitel-Amor2014} because the vorticity statistics are not reported.} 
   \label{fig:err_vs_Retau_zpg_turb}
\end{figure}

As shown in Fig.~\ref{fig:err_vs_Retau_zpg_turb}, for the four considered methods, i.e. the two mean-vorticity-based methods, the mean-shear method, and the local-reconstruction method, the relative errors between the computed and the classical definitions of the boundary-layer thickness are plotted versus the friction Reynolds number. The mean-shear method is normalized in inner units and a threshold constant of $1\times 10^{-3}$ is adopted following Vinuesa et al. \cite{Vinuesa2016}. It is observed that the two methods that rely on empirical threshold constants, i.e. the mean-shear method and the $-y\Omega_z$ threshold method, generate large errors $(\sim 10\%)$. Meanwhile, the local-reconstruction method and the $\widetilde{U}_\infty$ method are considerably more accurate with negligible errors for all considered Reynolds numbers.

\subsection{\label{sec:airfoils} Qualitative model assessment for the flow over airfoils}

Since there is no established definition for the boundary-layer thickness $\delta$ in complex flows, it is challenging to quantitatively assess the performance of various methods. On the other hand, in many scenarios, a qualitative examination is important and sufficient for turbulence model development.

\begin{figure}
  \centering
  \includegraphics[width=0.5\linewidth]{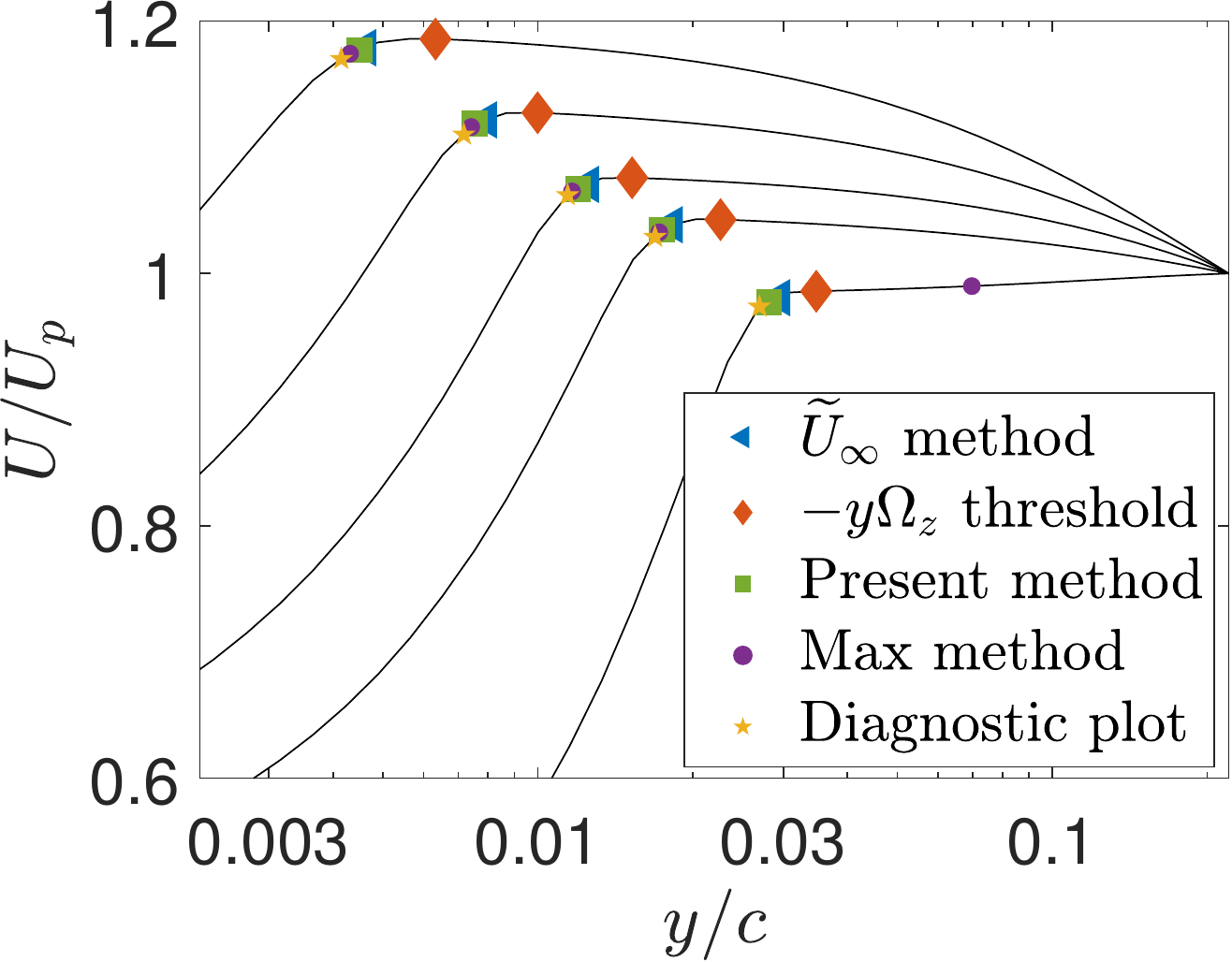}
  \caption{Distributions of the mean streamwise velocity (black solid lines) versus the wall-normal coordinate $y$ for various profiles from the suction side of a NACA 4412 at AoA $=5^{\circ}$ and $Re_c=10^6$ detailed in Table~\ref{tab:database}. These wall-normal profiles originate from the airfoil surface at the streamwise stations $x/c=0.20, 0.37, 0.55, 0.72, 0.90$ (from top-left to bottom-right). The estimates of the boundary-layer edges (and the corresponding edge velocities) are plotted with symbols as indicated in the legend.}
  \label{fig:del99_comp_demoD}
\end{figure}

\begin{figure}
  \centering
  \includegraphics[width=0.5\linewidth]{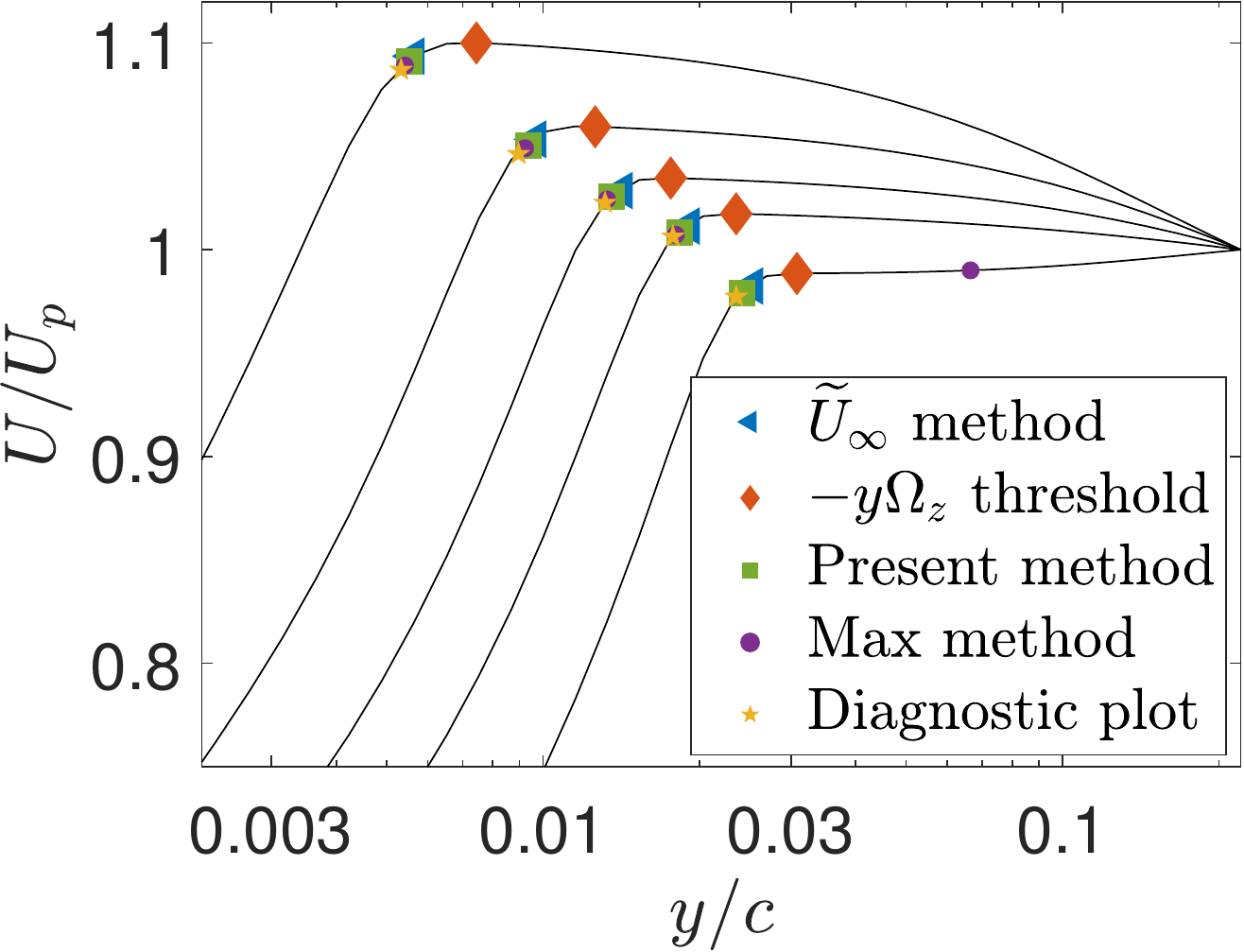}
  \caption{Distributions of the mean streamwise velocity (black solid lines) versus the wall-normal coordinate $y$ for various profiles from a symmetric NACA 0012 at AoA $=0^{\circ}$ and $Re_c=400,000$ detailed in Table~\ref{tab:database}. These wall-normal profiles originate from the airfoil surface at the streamwise stations $x/c=0.21, 0.37, 0.55, 0.72, 0.90$ (from top-left to bottom-right).
  The estimates of the boundary-layer edges (and the corresponding edge velocities) are plotted with symbols as indicated in the legend.}
  \label{fig:del99_comp_demoF}
\end{figure}

Considering the flow over the suction side of a NACA 4412 airfoil at the AoA $=5^{\circ}$ and the Reynolds number based on the chord length $Re_c=10^6$, Fig.~\ref{fig:del99_comp_demoD} shows the estimates for $\delta_{99}$ from the five previously discussed methods at five different streamwise stations. Note that the results from the $\widetilde{U}_\infty$ method, the diagnostic-plot method, and the local-reconstruction method are mutually consistent. The results from the ``max method," as the special case of the mean-shear threshold method with the threshold constant of zero, are also consistent, except at the most downstream station where the velocity profile is monotonically increasing. Meanwhile, the $-y\Omega_z$ threshold method systematically predicts a boundary-layer thickness larger than the others, implying that the threshold suggested by Uzun \& Malik \cite{Uzun2020} is not optimal for this case. Similar conclusions also hold when deploying these methods to the flow over a symmetric NACA 0012 airfoil at AoA $=0^{\circ}$ and Reynolds number $Re_c=400,000$, as shown in Fig.~\ref{fig:del99_comp_demoF}. 

\begin{figure}
  \centering
  \includegraphics[width=0.5\linewidth]{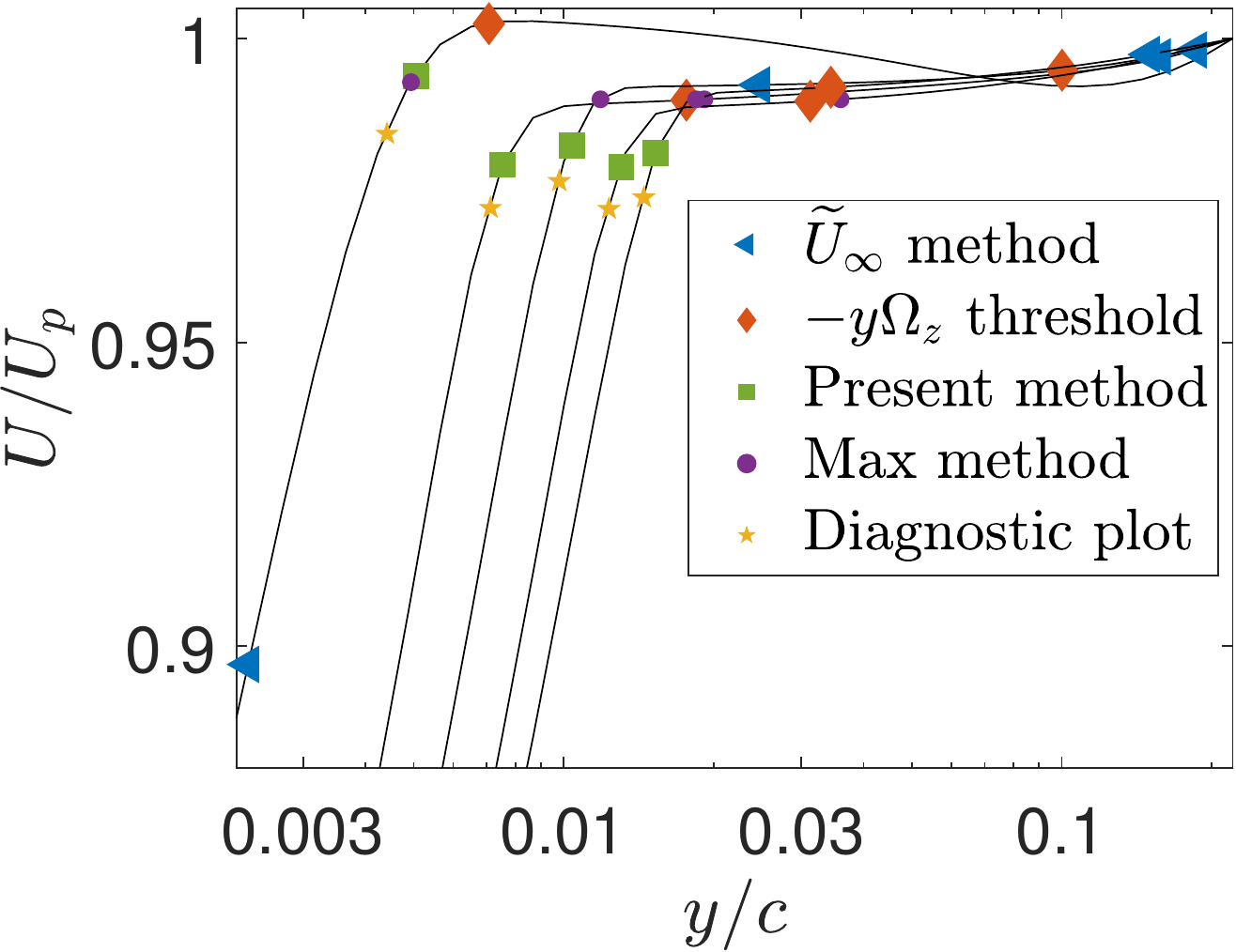}
  \caption{Distributions of the mean streamwise velocity (black solid lines) versus the wall-normal coordinate $y$ for various profiles from of a NACA 4412 at AoA $=5^{\circ}$ and $Re_c=10^6$. This is the same as Fig.~\ref{fig:del99_comp_demoD} except that included profiles are from the pressure side of the airfoil (see Table~\ref{tab:database} for details) and the wall-normal profiles originate from the airfoil surface at the streamwise stations $x/c=0.21, 0.37, 0.54, 0.72, 0.89$ (from left to right). The estimates of the boundary-layer edges (and the corresponding edge velocities) are plotted with symbols as indicated in the legend.}
  \label{fig:del99_comp_demoE}
\end{figure}

On the contrary, for the flow over the pressure side of the NACA 4412 airfoil at AoA $=5^{\circ}$ and Reynolds number $Re_c=10^6$, the conclusions are quite different. As shown in Fig.~\ref{fig:del99_comp_demoE}, only the results from the diagnostic plot method and the local-reconstruction method are equivalent. Qualitatively, these methods seem to identify the boundary-layer edge in a manner that is consistent both between the five considered profiles and the predictions in Fig.~\ref{fig:del99_comp_demoD} and Fig.~\ref{fig:del99_comp_demoF}. On the other hand, the $\widetilde{U}_\infty$ method severely under-predicts $\delta_{99}$ at the most upstream station and over-predicts $\delta_{99}$ at other stations. The $-y\Omega_z$ threshold method also overestimates $\delta_{99}$. As expected, the max method is unreliable, since several of the profiles are monotonically increasing and their maxima lie at $y_p$, a length scale that is not intrinsically connected to the boundary-layer thickness.

\begin{figure}
\begin{subfigure}{.5\textwidth}
  \centering
  \includegraphics[width=1\linewidth]{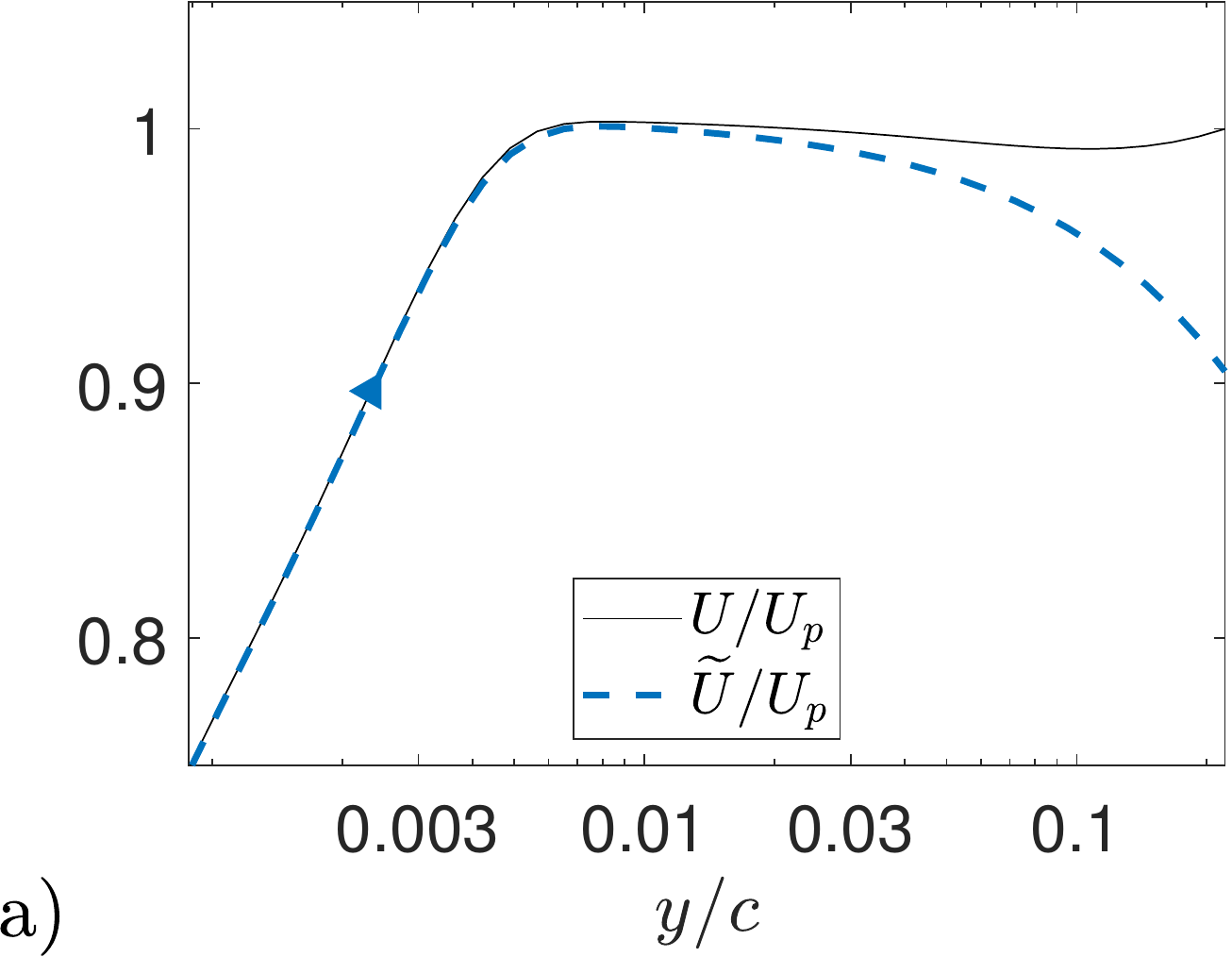}
  \captionlistentry{}
  \label{fig:U_tilde_vs_y_demoB}
\end{subfigure}%
\begin{subfigure}{.5\textwidth}
  \centering
  \includegraphics[width=1\linewidth]{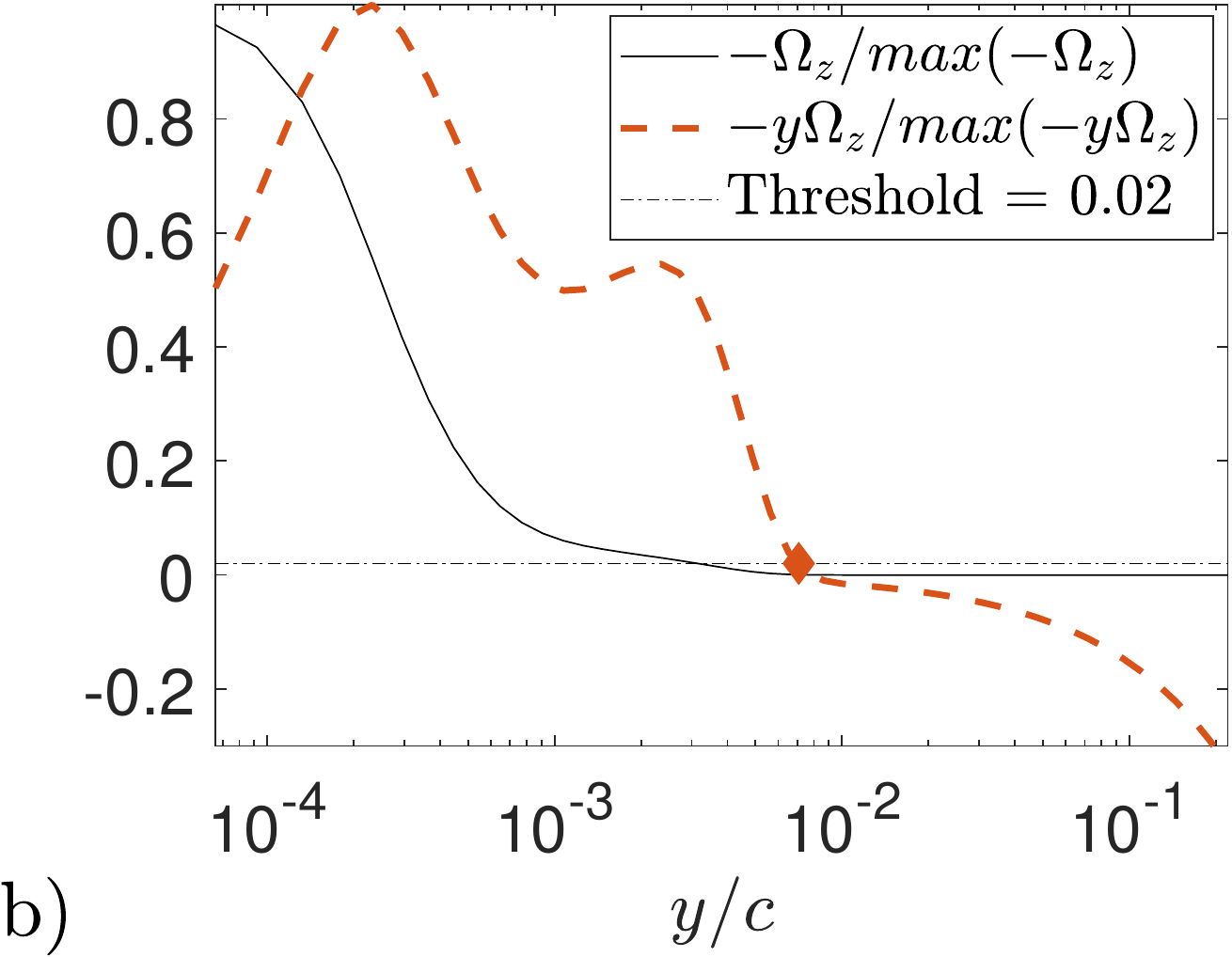}
  \captionlistentry{}
  \label{fig:Omega_z_vs_y_demoB}
\end{subfigure}%
 \hfill
\begin{subfigure}{.5\textwidth}
  \centering
  \includegraphics[width=1\linewidth]{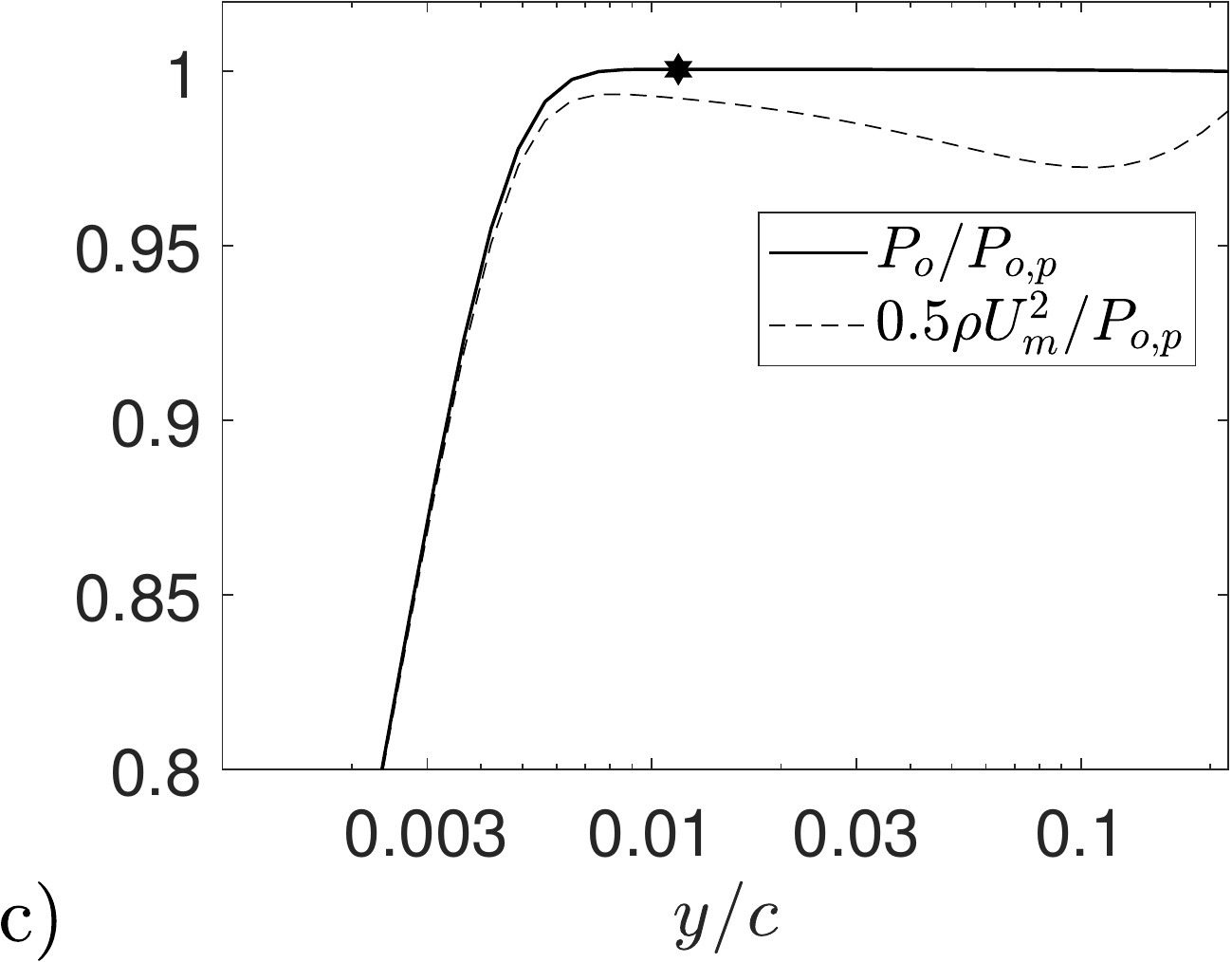}
  \captionlistentry{}
  \label{fig:Po_vs_y_demoB}
\end{subfigure}%
\begin{subfigure}{.5\textwidth}
  \centering
  \includegraphics[width=1\linewidth]{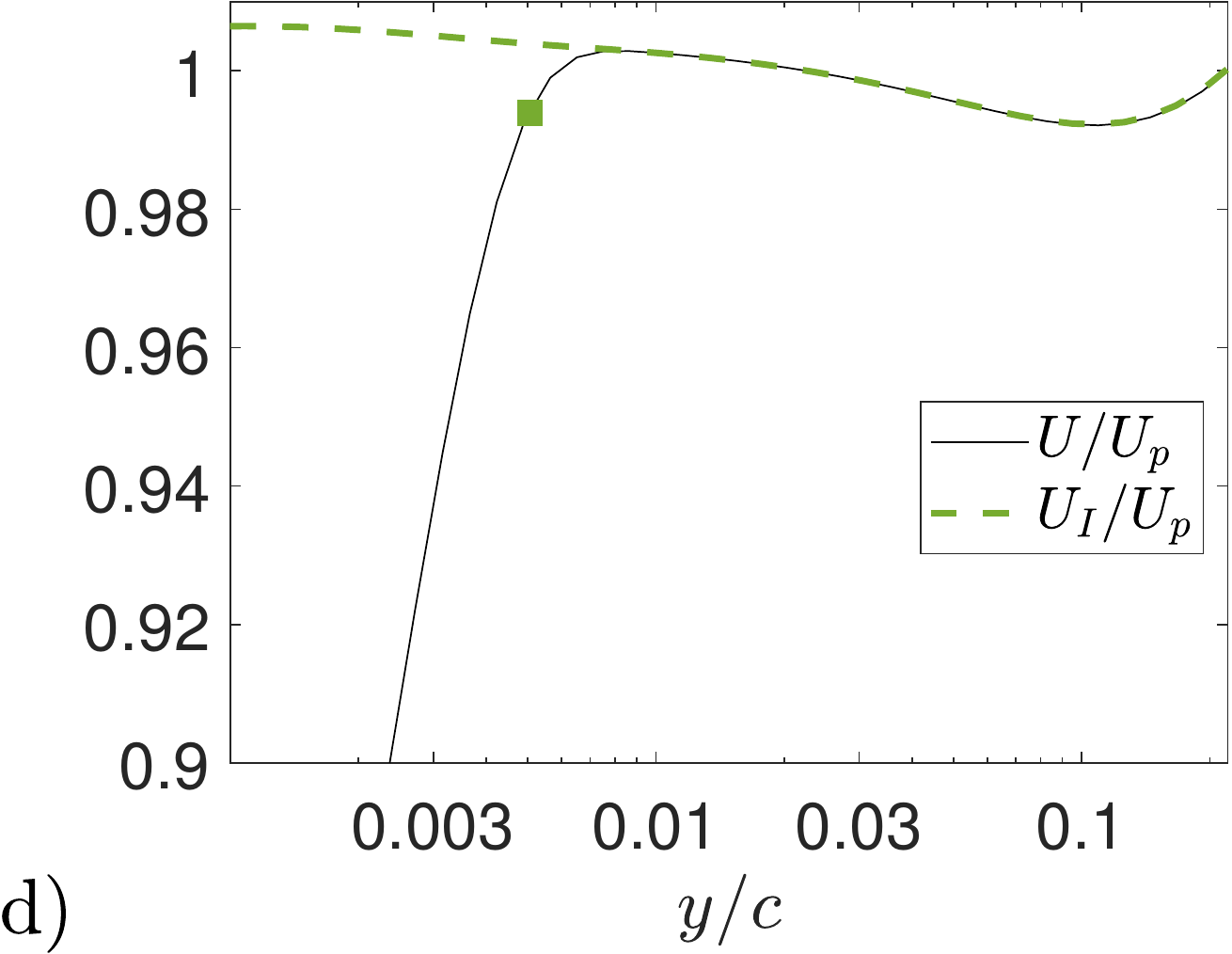}
  \captionlistentry{}
  \label{fig:UI_vs_y_demoB}
\end{subfigure}%
\caption{Distributions of the mean streamwise velocity $U$ and the generalized velocity $\widetilde{U}$ (a), the mean vorticity $\Omega_z$, the $-y\Omega_z$ boundary layer sensor, and the threshold constant (b), the stagnation pressure $P_o$ and the dynamic pressure $0.5 \rho U_m^2$ (c), and the mean velocity $U$ and the locally reconstructed inviscid solution $U_I$ (d) versus the wall-normal coordinate $y$ for the pressure side of a NACA 4412 airfoil at AoA $=5^{\circ}$ and $Re_c=10^6$ detailed in Table~\ref{tab:database}. All data is from the streamwise location $x/c = 0.21$. The estimates of the boundary-layer edges from the respective methods (a,b,d) and maximum stagnation pressure (c) are plotted with symbols. The subscript $p$ denotes quantities evaluated at the furthest point from the wall in the wall-normal profile.}
\label{fig:demoB_4plots}
\end{figure}

To investigate the reason for the failure of the mean-vorticity-based methods on the pressure side of the airfoil, the flow is further analyzed at the most upstream (i.e. $x/c=0.21$) and the most downstream (i.e. $x/c=0.89$)  stations from Fig.~\ref{fig:del99_comp_demoE}. At the station $x/c=0.21$, as shown in Fig.~\ref{fig:U_tilde_vs_y_demoB}, the generalized velocity does not achieve a constant value as $y \rightarrow \infty$. In fact, $\widetilde{U}_p$ is about $10\%$ less than the peak value of $\widetilde{U}$, which leads to an under-prediction of $\delta_{99}$. For the $-y\Omega_z$ threshold method, as shown in Fig.~\ref{fig:Omega_z_vs_y_demoB}, the distribution of the $-y\Omega_z$ profile is as expected. The over-prediction of $\delta_{99}$ is mainly attributed to the choice of the threshold constant. Despite the fact that the mean vorticity does not vanish as $y\rightarrow \infty$ as shown in Fig.~\ref{fig:Omega_z_vs_y_demoB}, it is observed that the stagnation pressure approaches a constant value outside the boundary layer, as shown in Fig.~\ref{fig:Po_vs_y_demoB}. As a result, the locally reconstructed ``invicid'' $U_I$ profile agrees well with the viscous solution $U$ in the freestream region and the divergence of these profiles indicates the location of the boundary-layer edge, see Fig.~\ref{fig:UI_vs_y_demoB}. On the other hand, the failure of the $\widetilde{U}_\infty$ method is mainly because of the sensitivity to the finite positive vorticity present outside the boundary layer. For the flow over an airfoil, it is expected that the vorticity will vanish outside the boundary layer. However, simulations of airfoils commonly use stretched stranded meshes, which are coarse in the streamwise and wall-normal directions near the boundary-layer edge. This can lead to relatively large errors in computing the spatial derivatives in the definition of the mean vorticity. Though the far-field vorticity observed in Fig.~\ref{fig:Omega_z_vs_y_demoB} may be spurious, it is a concern that the mean-vorticity-based methods are sensitive to this error.

\begin{figure}
\begin{subfigure}{.5\textwidth}
  \centering
  \includegraphics[width=1\linewidth]{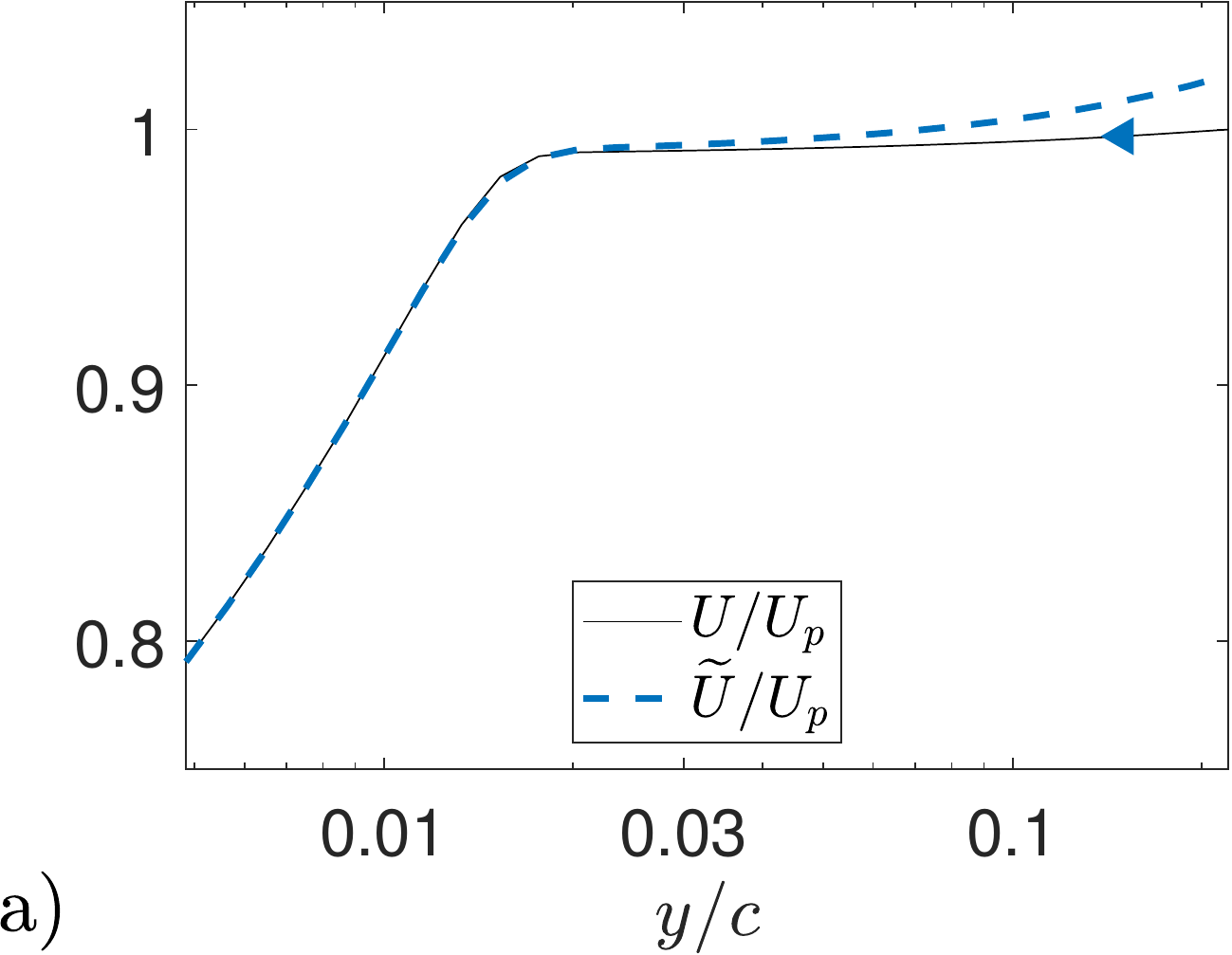}
  \captionlistentry{}
  \label{fig:U_tilde_vs_y_demoG}
\end{subfigure}%
\begin{subfigure}{.5\textwidth}
  \centering
  \includegraphics[width=1\linewidth]{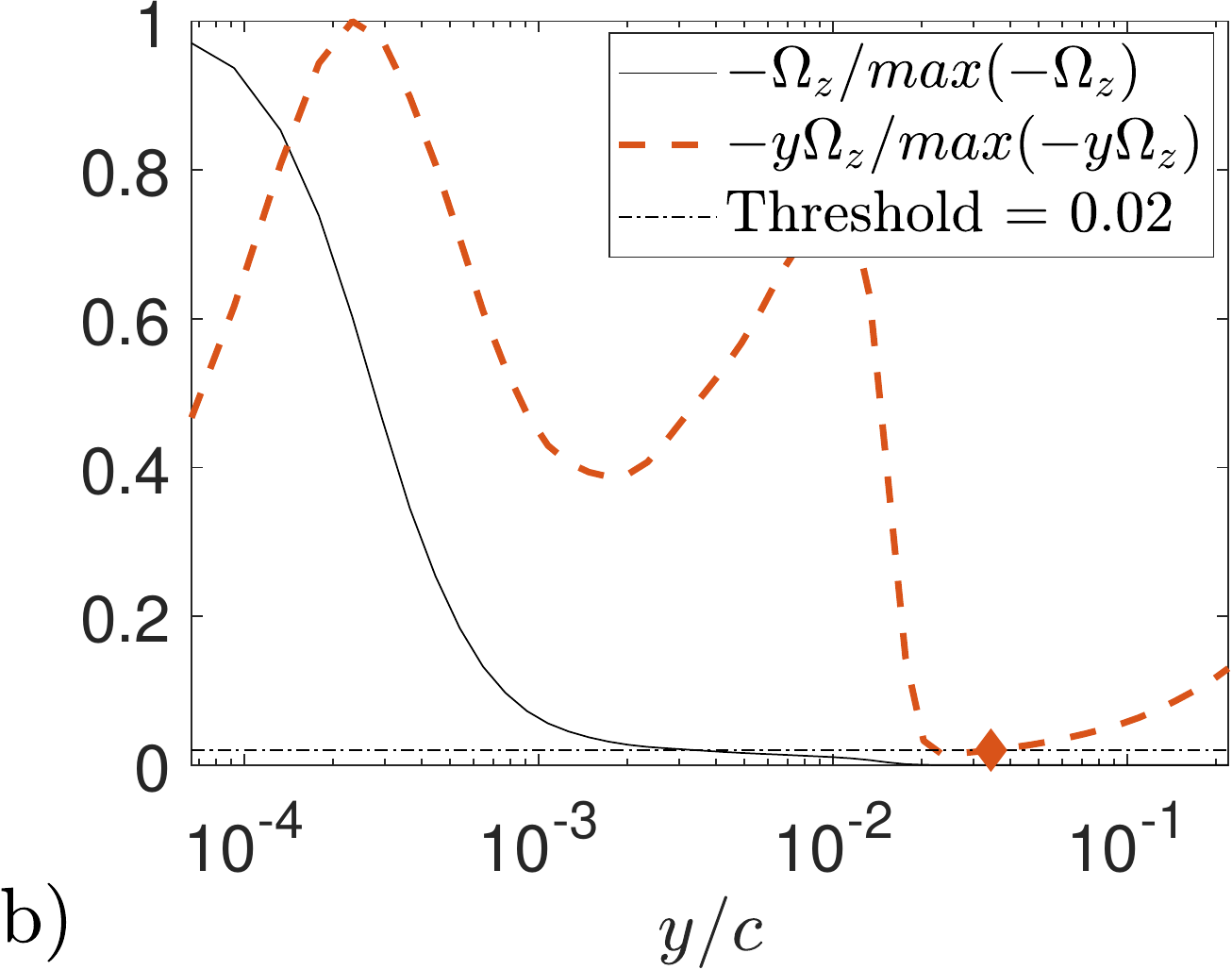}
  \captionlistentry{}
  \label{fig:Omega_z_vs_y_demoG}
\end{subfigure}%
\hfill
\begin{subfigure}{.5\textwidth}
  \centering
  \includegraphics[width=1\linewidth]{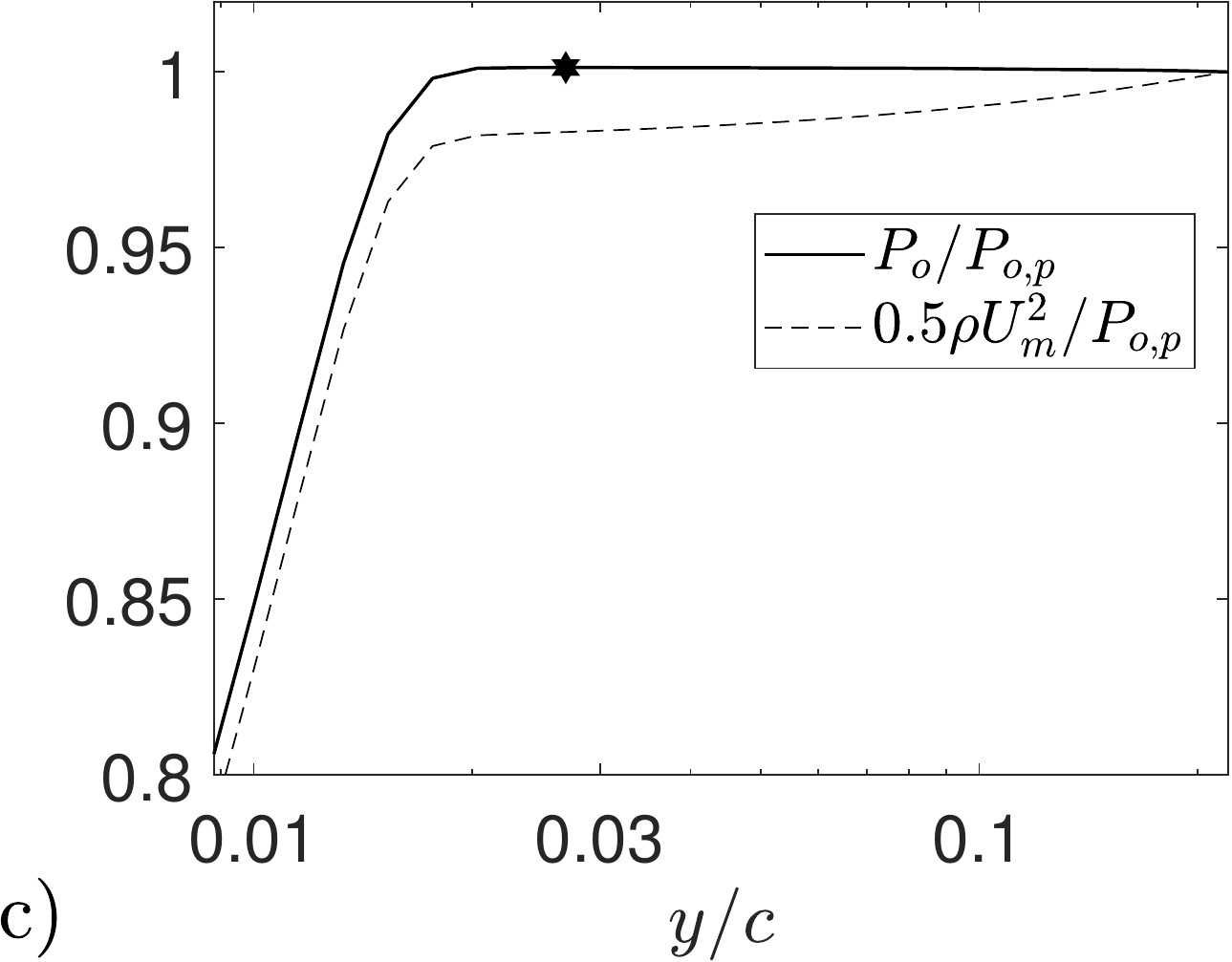}
  \captionlistentry{}
  \label{fig:Po_vs_y_demoG}
\end{subfigure}%
\begin{subfigure}{.5\textwidth}
  \centering
  \includegraphics[width=1\linewidth]{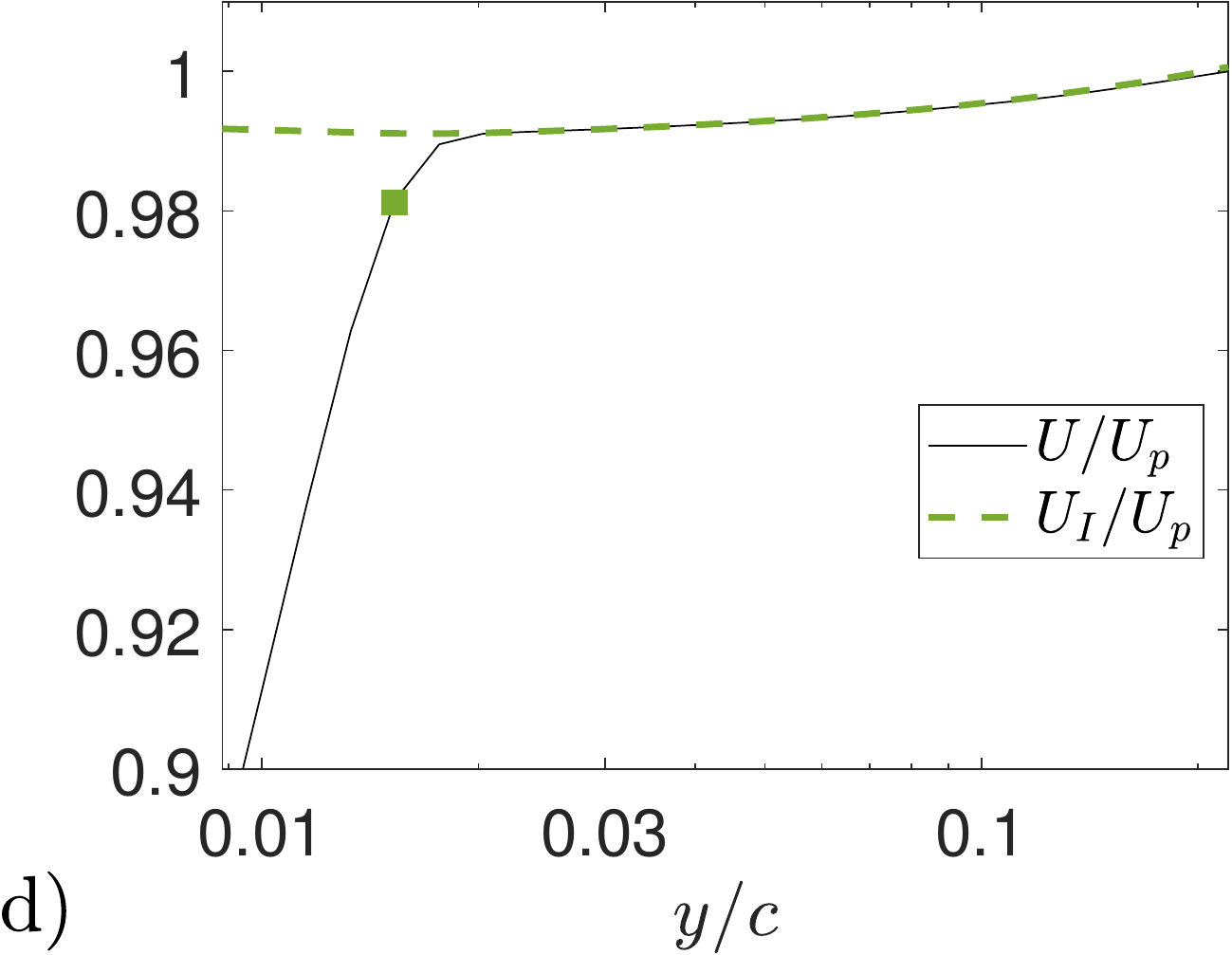}
  \captionlistentry{}
  \label{fig:UI_vs_y_demoG}
\end{subfigure}%
\caption{This is the same as Fig.~\ref{fig:demoB_4plots} except that the data is taken from wall-normal profiles originating from the airfoil surface at the streamwise station $x/c = 0.89$. The estimates of the boundary-layer edges from the respective methods (a,b,d) and maximum stagnation pressure (c) are plotted with symbols.}
\end{figure}

At the downstream station $x/c=0.89$, both of the mean-vorticity-based methods are behaving anomalously. As shown in Fig.~\ref{fig:U_tilde_vs_y_demoG}, the $\widetilde{U}$ profile fails to achieve a constant value as $y \rightarrow y_p$. Also, the mean vorticity outside the boundary layer is sufficiently large such that, as shown in Fig.~\ref{fig:Omega_z_vs_y_demoG}, $-y\Omega_z$ increases as $y \rightarrow y_p$. Such a distribution is in conflict with the assumption of the $-y\Omega_z$ threshold method that the boundary-layer edge is the furthest point from the wall at which $-y \Omega_z / max(-y \Omega_z) < 0.02$. Also, there is no rigorous justification that this threshold value can be achieved if the far-field vorticity is sufficiently large. As remarked above, although the presence of vorticity outside the boundary layer is likely spurious, it presents a robustness concern for the mean-vorticity-based methods. On the other hand, the local-reconstruction method performs well when deployed in this case. As shown in Fig.~\ref{fig:Po_vs_y_demoG} and Fig.~\ref{fig:UI_vs_y_demoG}, the stagnation pressure achieves a constant asymptote, and the boundary-layer edge is reasonably detected.

\subsection{\label{sec:visc_effects} Analysis of a pressure gradient boundary layer and far-field viscous effects}
In this section, the various methods for computing the boundary-layer thickness are applied to a flat-plate APGBL from the well-resolved simulation database of Bobke et al. \cite{Bobke2017}. Although viscous effects are typically negligible for $y>\delta$ in an APGBL, in this database, vorticity does not vanish completely and the stagnation pressure also does not asymptotically approach a constant value in the far-field.

\begin{figure}
\begin{subfigure}{.5\textwidth}
  \centering
  \includegraphics[width=1\linewidth]{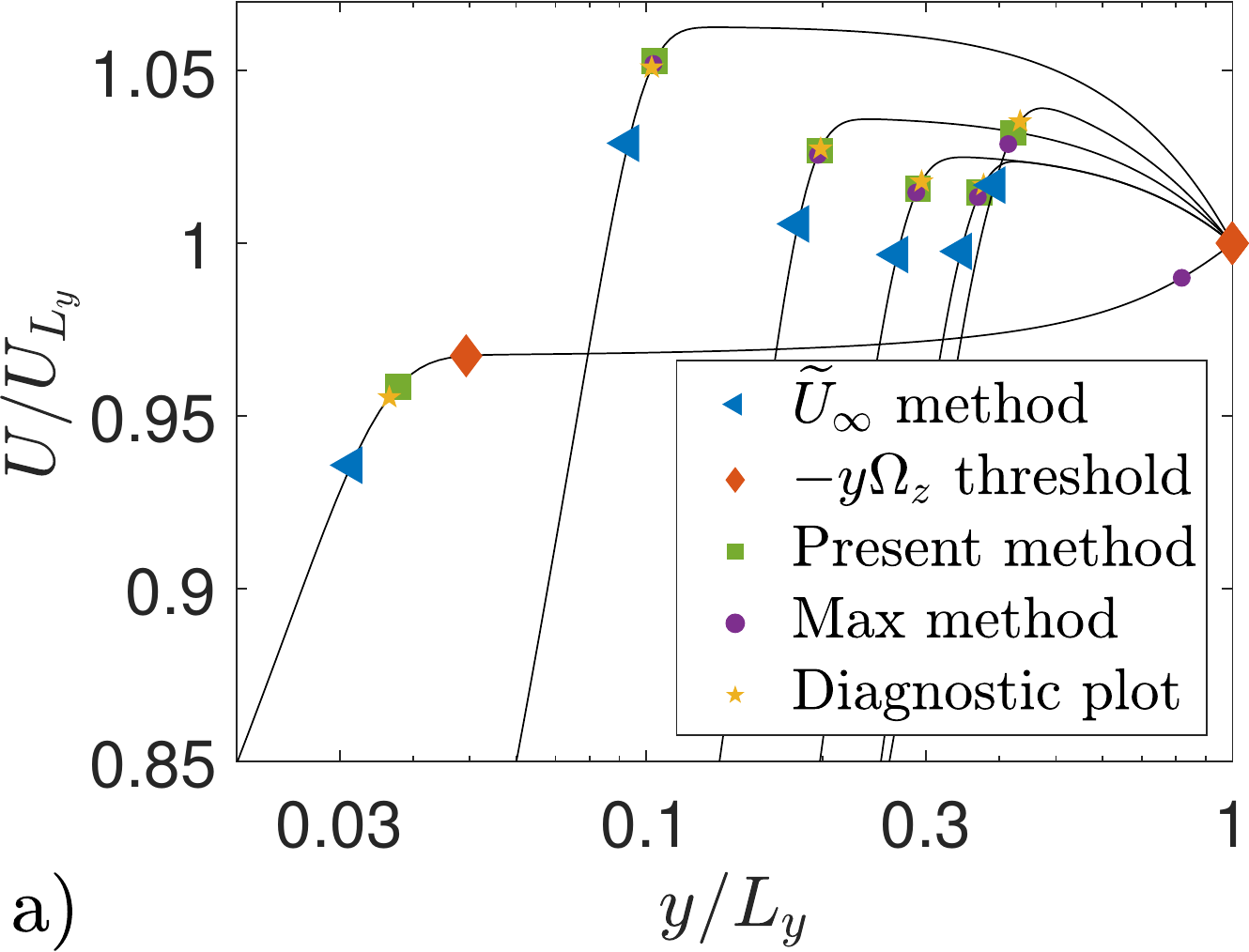}
  \captionlistentry{}
  \label{fig:del99_comp_demoH}
\end{subfigure}%
\begin{subfigure}{.5\textwidth}
  \centering
  \includegraphics[width=1\linewidth]{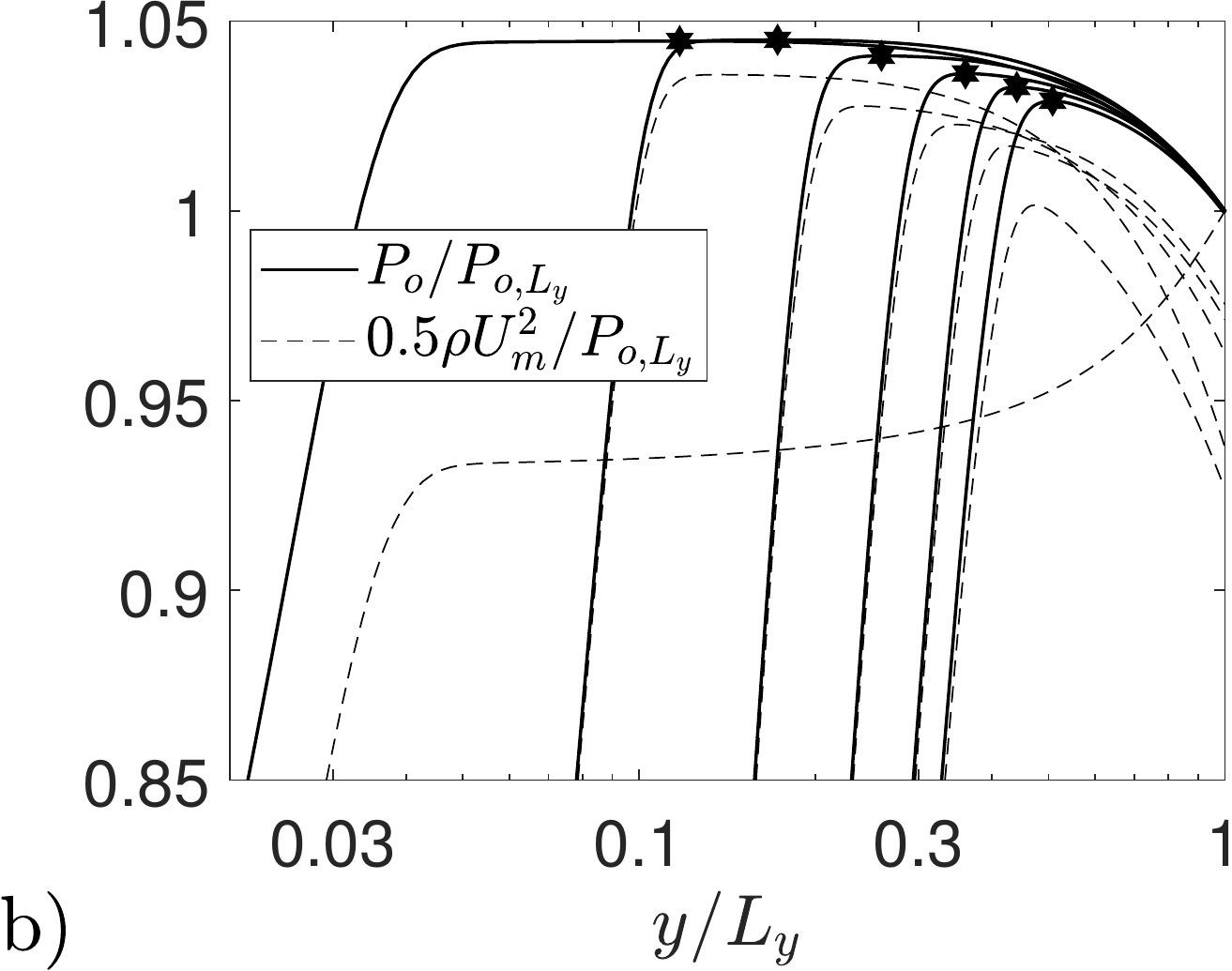}
  \captionlistentry{}
  \label{fig:Po_vs_y_demoH}
\end{subfigure}%
\caption{Distributions of the mean streamwise velocity (solid lines) in panel (a), and the stagnation (solid lines) and dynamic pressure (dashed lines) in panel (b) versus the wall-normal coordinate $y$ for various streamwise stations in a turbulent APGBL \cite{Bobke2017} with details specified in Table~\ref{tab:database}. These wall-normal profiles originate from the airfoil surface at the streamwise stations $x/L_x = 0.09, 0.27, 0.44, 0.62, 0.79, 0.97$ (from left to right), where $L_x$ denotes the length of the plate. The subscript $L_y$ refers to quantities evaluated on the domain boundary at $y=L_y$. In panel (a), the estimates of the boundary-layer edges (and the corresponding edge velocities) are plotted with symbols as indicated in the legend. In panel (b), the maxima of the stagnation pressure profiles are also indicated with symbols.}
\end{figure}

Estimates for the boundary-layer thickness are shown in Fig.~\ref{fig:del99_comp_demoH} for an APGBL \cite{Bobke2017} (referred to as case $m18$ therein). The diagnostic-plot and the local-reconstruction methods provide consistent predictions of the boundary-layer thickness for all six streamwise stations considered. The max method also provides a consistent estimate except at the streamwise station with a monotonically increasing profile. On the other hand, the $\widetilde{U}_\infty$ method systematically under-predicts $\delta_{99}$, while the $-y\Omega_z$ threshold method erroneously computes $\delta_{99}\approx L_y$ with significant errors for most cases, where $L_y$ is the extent of the domain in the wall-normal direction. In Fig.~\ref{fig:Po_vs_y_demoH}, distributions of the stagnation pressure and the dynamic pressure are plotted versus the wall-normal coordinate. The profiles achieve the global maxima in the interior of the domain and the local minima at $y=0$ and $y=L_y$. The local-reconstruction method, however, can handle such scenarios robustly.

\begin{figure}
\begin{subfigure}{.5\textwidth}
  \centering
  \includegraphics[width=1\linewidth]{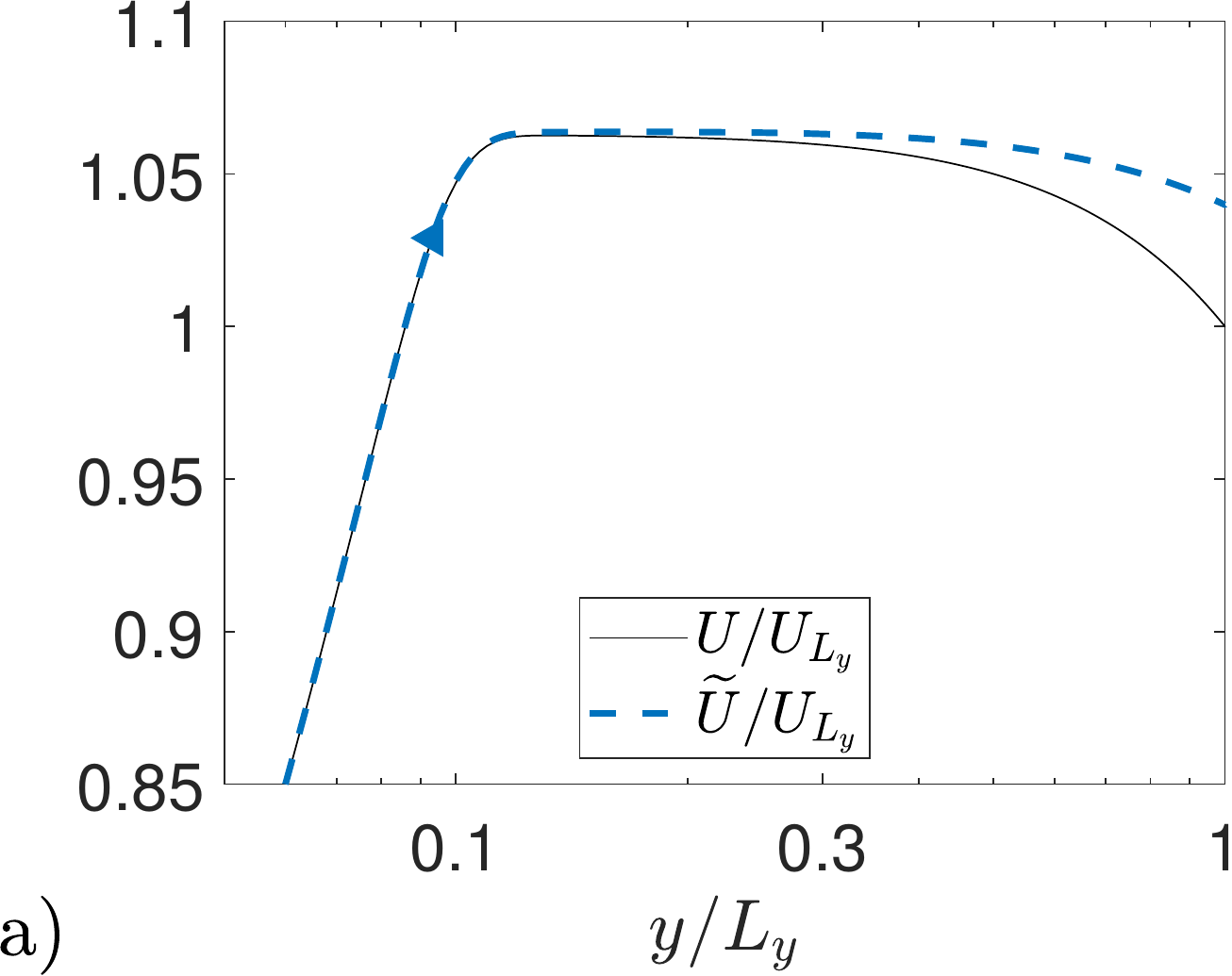}
  \captionlistentry{}
  \label{fig:U_tilde_vs_y_demoI}
\end{subfigure}%
\begin{subfigure}{.5\textwidth}
  \centering
  \includegraphics[width=1\linewidth]{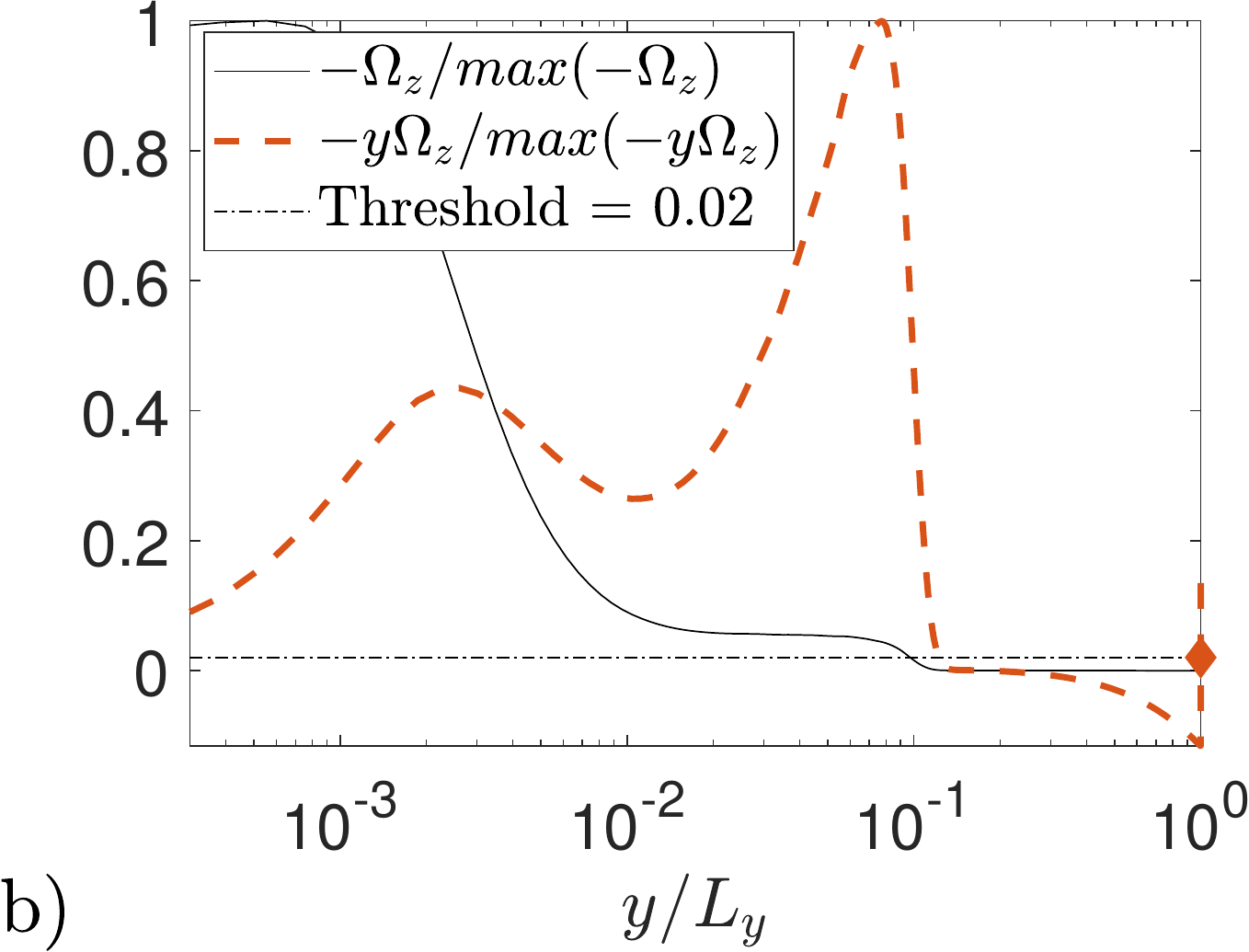}
  \captionlistentry{}
  \label{fig:Omega_z_vs_y_demoI}
\end{subfigure}%
\hfill
\begin{subfigure}{.5\textwidth}
  \centering
  \includegraphics[width=1\linewidth]{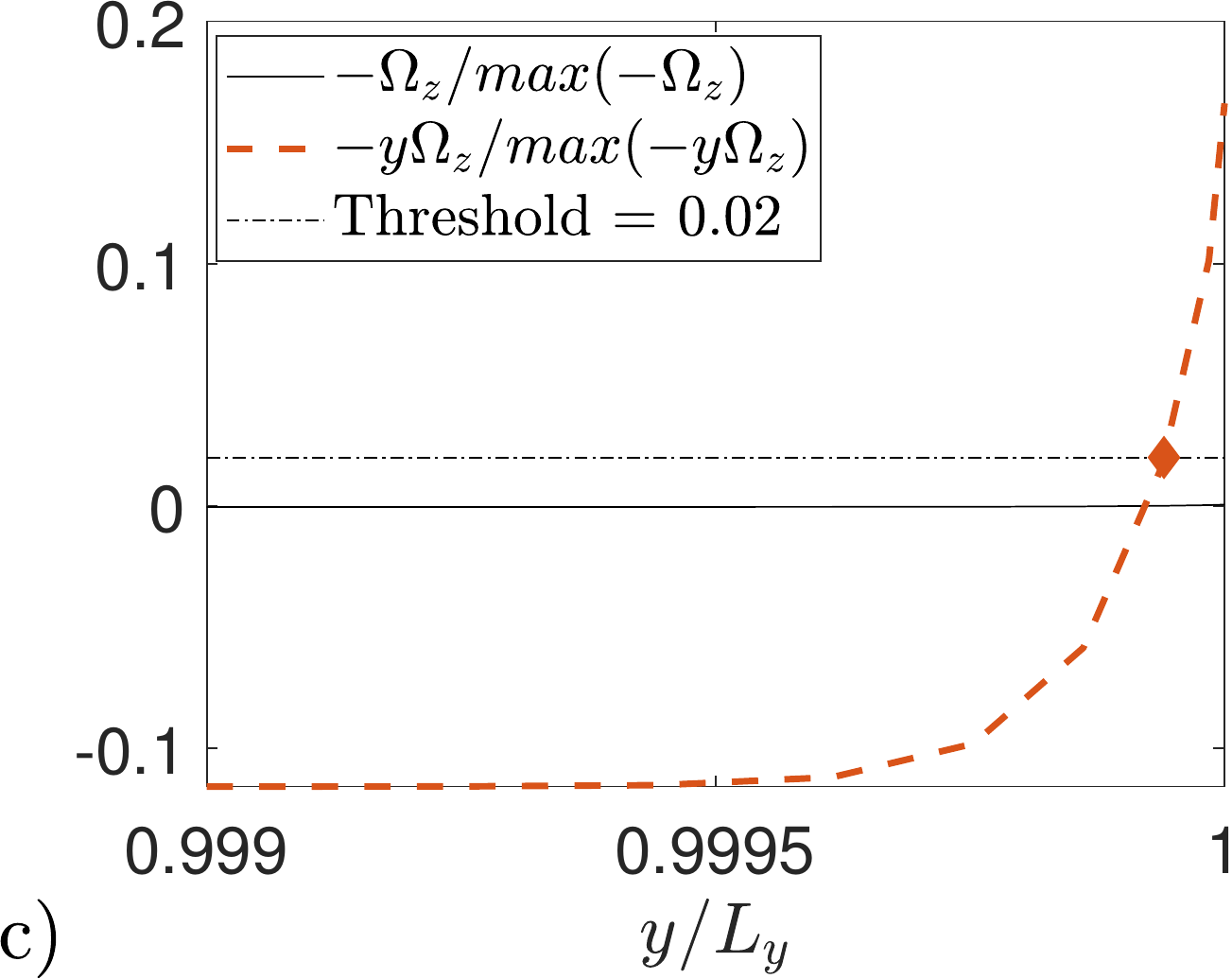}
  \captionlistentry{}
  \label{fig:Omega_z_zoom_demoI}
\end{subfigure}%
\begin{subfigure}{.5\textwidth}
  \centering
  \includegraphics[width=1\linewidth]{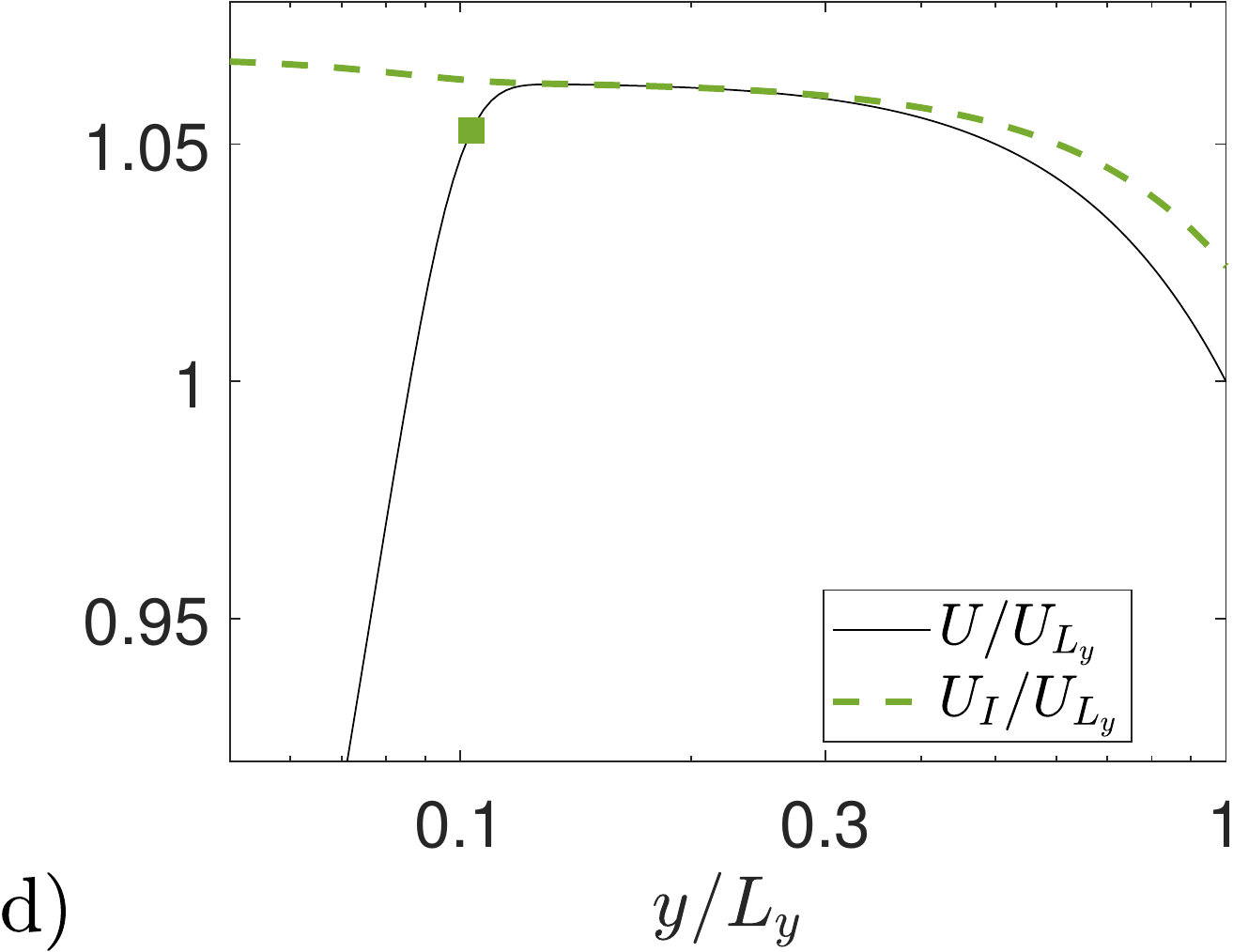}
  \captionlistentry{}
  \label{fig:UI_vs_y_demoI}
\end{subfigure}%
\caption{Distributions of the mean streamwise velocity $U$ and the generalized velocity $\widetilde{U}$ (a), the mean vorticity $\Omega_z$, the $-y\Omega_z$ boundary layer sensor, and the threshold constant (b), the mean velocity $U$ and the locally reconstructed inviscid solution $U_I$ (d) versus the wall-normal coordinate $y$ for the turbulent APGBL \cite{Bobke2017} detailed in Table~\ref{tab:database}. Panel (b) is reproduced in panel (c) on zoomed linear axes. All data is taken from the streamwise station at $x/L_x = 0.27$. The estimates of the boundary-layer edges (and the corresponding edge velocities) from the respective methods are plotted with symbols.}
\end{figure}

To explain the failure of the mean-vorticity-based methods in Fig.~\ref{fig:del99_comp_demoH}, the second streamwise station at $x/L_x = 0.27$ is examined in detail. As shown in Fig.~\ref{fig:U_tilde_vs_y_demoI}, the generalized velocity $\widetilde{U}$ continuously decays rather than achieving a constant asymptote as $y \rightarrow L_y$, which is the reason for the systematic underestimate of $\delta_{99}$ by the $\widetilde{U}_\infty$ method in Fig.~\ref{fig:del99_comp_demoH}. In Fig.~\ref{fig:Omega_z_vs_y_demoI}, the $-y\Omega_z$ profile is shown to cross the threshold three times, whereas this threshold is typically assumed to only be crossed twice \cite{Uzun2020}. As indicated by Fig.~\ref{fig:Omega_z_zoom_demoI}, it is the third crossing that leads to the wrong identification of $\delta_{99} \approx L_y$. As shown in Fig.~\ref{fig:UI_vs_y_demoI}, the locally-reconstructed inviscid solution $U_I$ agrees well with the velocity $U$ profile in the middle area of the domain. Moreover, as expected, the $U_I$ profile departs from the $U$ profile near the wall and in the vicinity of the top boundary, due to the fact that viscous effects are significant in both of these regions. Nonetheless, the boundary-layer edge is well identified by the local-reconstruction method for this critical case.

\subsection{\label{sec:internal_flows} Consistency in channel and pipe flows}

For a symmetric, fully developed channel or pipe flow, the boundary-layer thickness is conventionally defined as the half-height of the geometry. However, most above mentioned existing methods for computing $\delta$ may not recover this behavior. Although a programmatic way of computing $\delta$ is hardly needed for a fully developed channel or pipe flow, in complex or developing internal flows, such as a serpentine duct with separation, the boundary-layer thickness is no longer trivially defined by the geometry. This motivates interrogating each method for computing $\delta$ to see if the expected behavior is recovered in a canonical internal flow, as this is a necessary condition for a reliable prediction in complex internal flows.

Specifically, the intermittency-threshold and diagnostic-plot-based methods assume that the intermittency and turbulence intensity, respectively, decay below a predefined threshold at the boundary-layer edge. However, there is no guarantee that these thresholds will be crossed at the centerline of a fully-developed channel or pipe flow. On the other hand, the fully-developed nature of channel and pipe flows implies that $\Omega_z = \pderi{U}{y}$ and $\widetilde{U} = U$ since $V=0$. This implies that the $-y\Omega_z$ threshold method and the mean-shear-threshold method correctly identify the centerline as the boundary-layer edge if the corresponding thresholds are properly set to zero. By construction, the $\widetilde{U}_\infty$ method fails in a fully developed channel or pipe flow due to the fact that the generalized velocity $\widetilde{U}$ continuously varies in the entire domain rather than asymptotically approaching a constant value. For the local-reconstruction method, since $(\partial P/\partial y)[y]=V[y]=0$, the peak stagnation pressure occurs at the centerline, implying that the locally reconstructed $U_I$ is equal to the centerline velocity $U$. Consequently, the predicted boundary-layer thickness $\delta_{100}$ corresponds to the half height of the channel as expected. Note that the above discussions apply to both turbulent and laminar flows in channels and pipes.

\subsection{Comparative evaluation} \label{sec:comparison}

To evaluate the robustness and the general applicability of the methods discussed above, the potentially undesirable features of each method are summarized in Table~\ref{tab:methods}. The columns of the table refer to the existing methods reviewed above and the proposed local-reconstruction method. The rows of the table are discussed in a line-by-line manner below. 

The first row indicates whether the method requires numerical integration, which may introduce numerical truncation error that scales with the spatial resolution of the data. The second row indicates whether the
method relies on the computation of spatial derivatives or standard deviations, as such operations are sensitive to noise. The third row indicates whether the method employs empirical coefficients or profiles. The fourth row indicates whether the method invokes an iterative procedure, and thus requires an initial guess for the boundary-layer thickness. Although an iterative method may converge in simple applications, convergence for general velocity profiles can not be guaranteed. The fifth row indicates whether the method is restricted to high-Reynolds-number turbulent boundary layers with a quiet freestream. The last row indicates whether the method requires ``additional data" besides the mean streamwise velocity profile $U[y]$. For instance, the intermittency-threshold method and the diagnostic-plot method require time-resolved data, which may be unavailable in experiments or RANS simulations. 

As shown in Table~\ref{tab:methods}, the various existing methods feature much more undesirable properties than the local-reconstruction method, indicating the lack of robustness and the general applicability. The local-reconstruction method is devoid of unfavorable characteristics with the exception of requiring ``additional data," i.e. the $V[y]$ and $P[y]$ profiles, which are, however, often reported.

\begin{table}[]
\begin{tabular}{l"c|c|c|c|c|c|c}
\thead{Method}           & \thead{$\widetilde{U}_\infty$ \\ method} & \thead{$-y\Omega_z$ \\ threshold} & \thead{Mean-shear \\ threshold} & \thead{Diagnostic \\ plot} & \thead{Composite \\ profiles}  & \thead{Intermittency \\ threshold} & \thead{Local- \\ reconstruction \\ method} \\ \thickhline
Integral-based           &$\times$&        &        &$\times$&        &        &          \\\hline
\makecell[cl]{Computes derivatives \\ or
standard deviations}     &$\times$&$\times$&$\times$&$\times$&        &$\times$&          \\\hline
Employs empirical data   &        &$\times$&$\times$&$\times$&$\times$&$\times$&          \\\hline
Requires iteration       &$\times$&        &$\times$&$\times$&$\times$&        &          \\\hline
\makecell[cl]{Requires high $Re$ and/or \\ a
quiet freestream}        &$\times$&$\times$&        &$\times$&$\times$&$\times$&         \\\hline
Requires additional
data$^*$                   &$\times$&$\times$&        &$\times$&        &$\times$&$\times$ \\\thickhline
Total                    &   5    &    4   &    3   &   6    &   3    &    4   &    1 
\end{tabular}
\caption{Potentially unfavorable features (indicated by the ``$\times$" symbol) for the existing methods discussed in section~\ref{sec:existing_methods}, Appendix~\ref{app:composite_profiles}, and Appendix~\ref{app:intermittency} and the local-reconstruction method developed in section~\ref{sec:new_method}. The total number of potentially unfavorable features of each method is tabulated. $^*$ ``additional data" refers to any statistics other than $U[y]$.}
\label{tab:methods}
\end{table}

\section{\label{sec:conclusion}Conclusions}

Although the boundary-layer thickness can be rigorously computed in ZPGBLs, for flows with curved geometries or other non-equilibrium effects, there exist no agreed-upon definitions for the boundary-layer thickness and edge velocity. In this work, the limitations of various existing methods for computing the boundary-layer thickness are analyzed and contrasted with the proposed local-reconstruction method. Without relying on any empirical parameters, the local-reconstruction method employs the Bernoulli equation to locally reconstruct the inviscid velocity profile from the pressure and wall-normal velocity profiles and identifies the boundary-layer edge by comparing the reconstructed inviscid velocity profile with the viscous velocity profile. When deployed to the laminar and turbulent ZPGBLs, quantitative assessment reveals that only the local-reconstruction method achieves negligible errors without parameter tuning. Moreover, the proposed method is readily applied to more complex flows, e.g. flows over airfoils and an adverse pressure gradient boundary layer with far-field vorticity, with a consistent predictive capability.


\begin{acknowledgments}
KG acknowledges support from the National Defense Science and Engineering Graduate Fellowship and the Stanford Graduate Fellowship. LF and PM are funded by NASA grant No. NNX15AU93A. We wish to acknowledge helpful discussions with Adrian Lozano Duran, Sanjeeb T Bose, Javier Urzay, Suhas Suresh Jain, and Ahmed Elnahhas. 
\end{acknowledgments}

\appendix

\section{Methods based on composite profiles} \label{app:composite_profiles}

The boundary-layer thickness can be computed by assuming that the mean streamwise velocity profile has a functional form that depends on $\delta$ and by fitting this function to the mean velocity data. For turbulent flows, the assumed functional form of the velocity profile is referred to as a composite profile \cite{Vinuesa2016} because typically the profile is composed of an inner law of the wall and an outer law of the wake, e.g. the Coles' wall-wake formula \cite{Coles1956}, given by Eq.~(\ref{eq:coles}). More elaborate composite profiles (e.g. \cite{Nickels2004}) have been proposed since, but in this section, Coles' law \cite{Coles1956} is considered for illustration purposes. Given the velocity profile $U[y]$ from an experiment or simulation, the parameters, i.e. $\kappa$, $B$, $\Pi$, and $\delta$ in Eq.~(\ref{eq:coles}), can be determined by a least-squares fitting of the composite profile to the mean velocity data. Depending on the regime, a subset of these parameters may be specified from well-established correlations to reduce the dimensionality of the data fitting.

Vinuesa et al. \cite{Vinuesa2016} use the composite profile proposed by Nickels \cite{Nickels2004} to compute the boundary-layer thickness for the flow over a 2D airfoil. It is found that an iterative method is required and that the fitted composite profile does not agree well with the input velocity profile when the Reynolds number is low or the pressure gradient is strong. In other words, the resulting estimate for $\delta$ is only accurate insofar as the composite profile is rigorously applicable to the flow being considered. 
Due to the lack of general applicability, the composite-profile-based methods are not investigated in this work.
 
\section{Method based on an intermittency threshold} \label{app:intermittency}

For a turbulent boundary layer with a quiescent freestream, the boundary-layer thickness can be determined using a sensor based on the intermittency, which is defined locally as the fraction of time that the flow is turbulent. Common criteria for classifying the local flow as turbulent or laminar are to define empirical thresholds for the instantaneous vorticity magnitude $|\omega| = | \vec{\nabla} \times \vec{u} |$ or the instantaneous enstrophy $\omega \cdot \omega$. For instance, Jimenez et al. \cite{Jimenez2010} label the local flow as turbulent when the instantaneous vorticity magnitude $|\omega|^+ \ge 5\times 10^{-3}$, and laminar otherwise, resulting in a binary value distribution at every instance in space and time. Subsequently, the distribution of the intermittency function $\gamma$ can be obtained by time averaging this binary field. At last, the boundary-layer thickness is defined as the wall-normal distance at which the intermittency drops below a second empirical threshold. As shown by Murlis et al. \cite{Murlis1982}, and Li and Schlatter \cite{Li2011}, an intermittency threshold constant of 0.3 correlates with $\delta_{99}$ well.

An obvious weakness of this approach is the reliance on two empirical threshold constants. In addition, this method may fail in external flows with strong freestream turbulence since the prescribed thresholds might never be crossed. Also, this method does not function for laminar and transitional flows since the intermittency is always zero in the laminar portion of boundary layer. 

Moreover, the data required to deploy this method may not always be available. For instance, the intermittency distribution is rarely measured in experiments \cite{Vinuesa2016}. In LES, since intermittency is not a mesh-convergent statistic, the estimate of the boundary-layer thickness depends on the mesh resolution even if the mean velocity profile is well resolved. Nonetheless, the intermittency threshold method has been successfully deployed in LES \cite{Li2011}. Also, this method is not applicable in RANS simulations, where the intermittency is typically not modeled.

\section{Methods based on assumed inviscid profiles} \label{app:assumed_inviscid_profile}

The boundary-layer thickness can be estimated by assuming a functional form for the velocity profile that applies for $y > \delta$. For example, Kim and Simon~\cite{Kim1991} study the spatially developing boundary layer over a curved plate and assume that the flow outside of the boundary layer recovers the inviscid solution, which can be characterized by the hyperbola
\begin{equation}
    U_{I,\mathrm{hyper}} = \frac{U_{I,\mathrm{hyper}}[y=0]}{1-y k},
\end{equation}
where $k$ is the signed curvature of the surface. (Note that this profile neglects the perturbation by the displacement thickness which will perturb the effective geometry shape seen by the outer flow.)
Kim and Simon~\cite{Kim1991} determine $U_{I,\mathrm{hyper}}[y=0]$ by a least squares regression of two velocity samples at wall-normal distances that are assumed to be outside of the boundary layer. To validate this method, the two data samples nearest to the end of profile $U_p[y=y_p]$ are chosen in this work. Finally, this profile is used to estimate the boundary-layer thickness according to Eq.~(\ref{eq:delta_n_new}). 

A simplified version of this approach assumes that, outside the boundary layer, $U_I$ approaches the linear asymptote,
\begin{equation}
    U_{I,\mathrm{linear}} = m y + U_{I,\mathrm{linear}}[y=0],
\end{equation}
where the slope $m$ and the intercept $U_{I,\mathrm{linear}}[y=0]$ can be determined by fitting two velocity samples in the same manner as above.

\begin{figure}
\begin{subfigure}{.5\textwidth}
  \centering
  \includegraphics[width=1\linewidth]{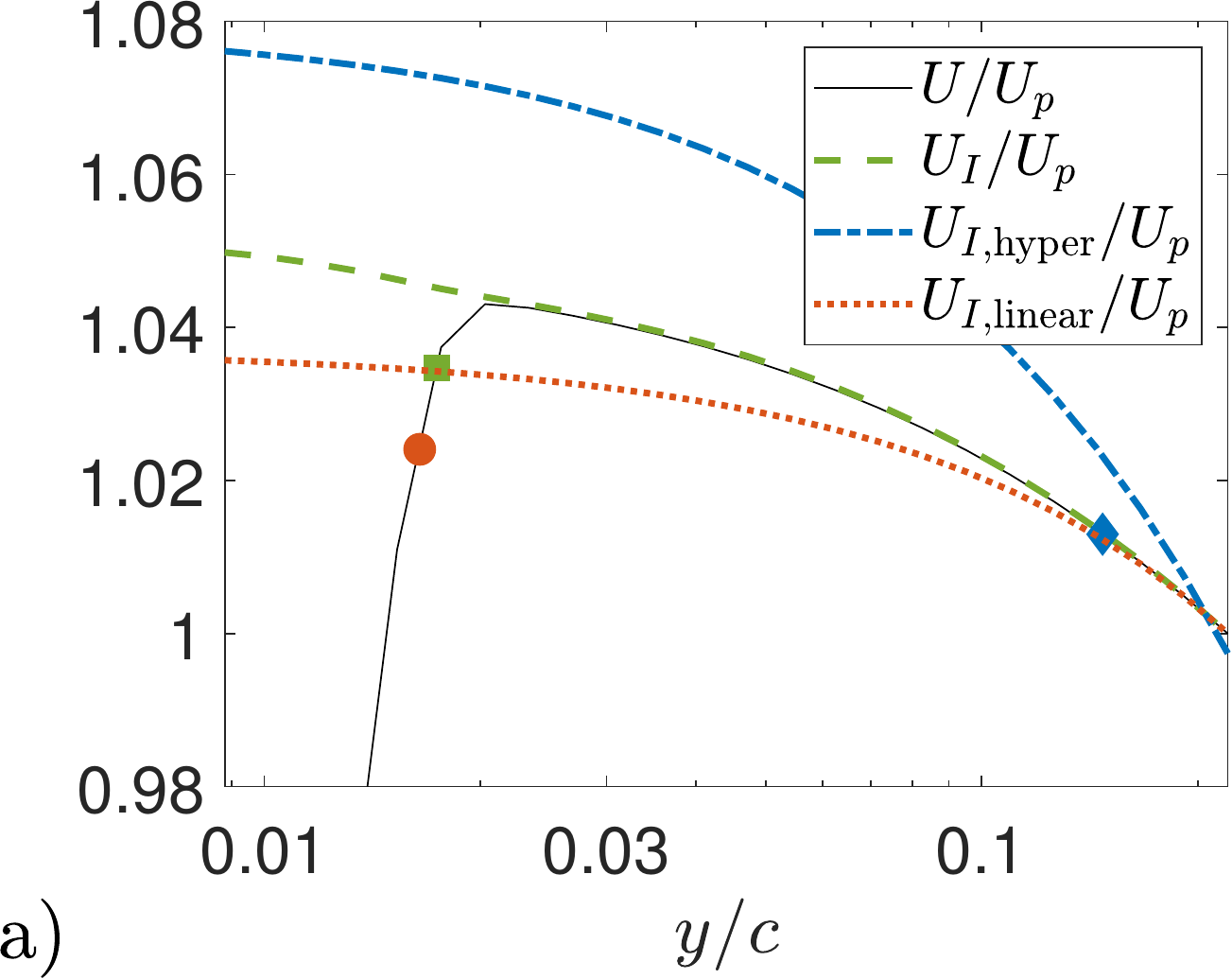}
  \captionlistentry{}
\end{subfigure}%
\begin{subfigure}{.5\textwidth}
  \centering
  \includegraphics[width=1\linewidth]{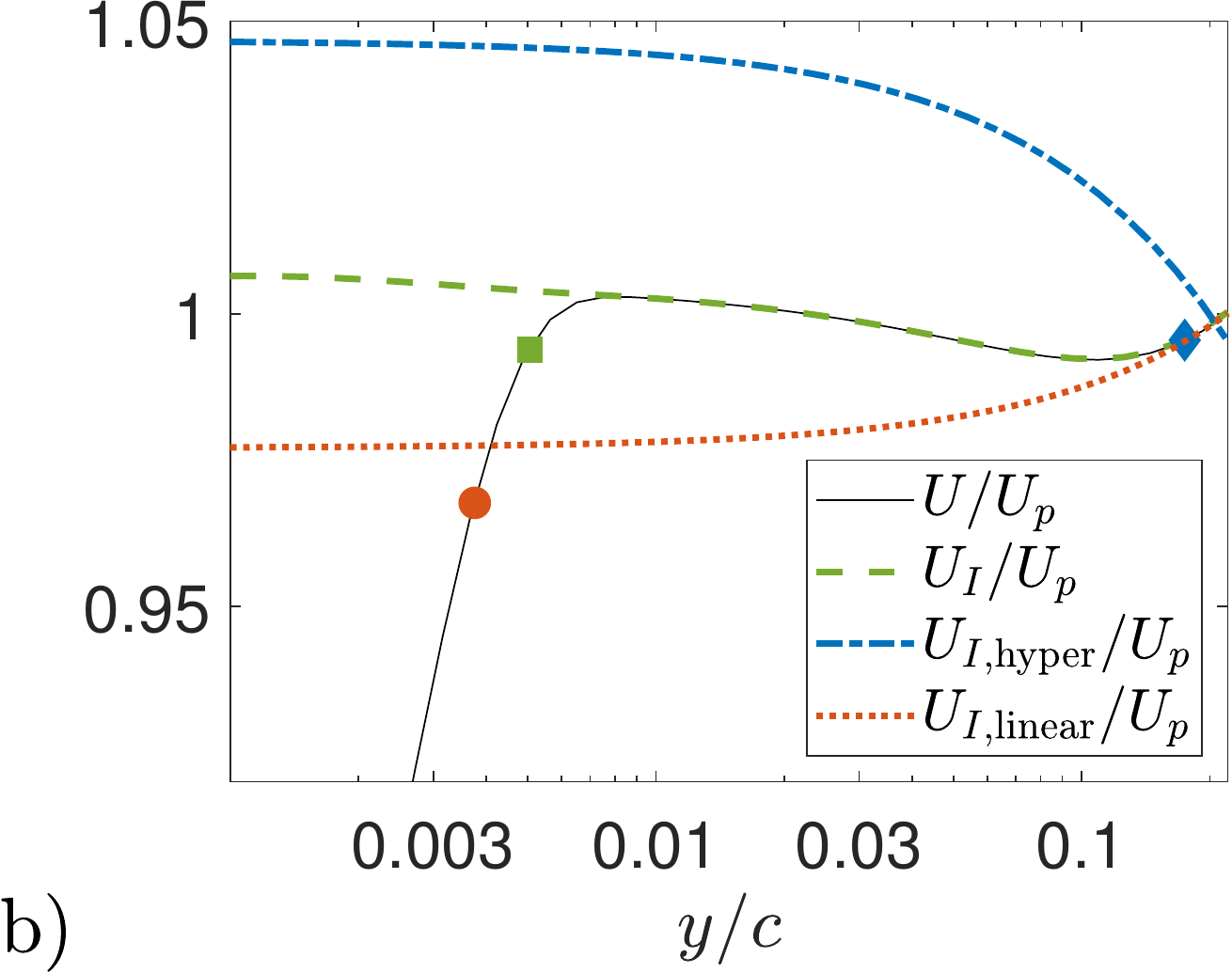}
  \captionlistentry{}
\end{subfigure}%
\caption{Distributions of the velocity profile $U$, the locally reconstructed inviscid velocity profile $U_I$, and the hyperbolic and linear approximations of $U_I$ versus the normalized wall-normal coordinate $y/c$. The boundary-layer thicknesses implied by Eq.~(\ref{eq:delta_n_new}) based on the three approximate inviscid solutions are indicated with squares, diamonds, and circles, respectively. All data are from a NACA 4412 airfoil at AoA $=5^{\circ}$ and $Re_c=10^6$ \cite{Vinuesa2018}. Specifically, the data are extracted from the suction side at the streamwise station $x/c = 0.72$ (a) and the pressure side at $x/c = 0.21$ (b).}
\label{fig:inviscid_profs}
\end{figure}

The reconstructed velocity profiles outside of the boundary layer from these two methods and the present method (as given by Eq.~(\ref{eq:U_I_bernoulli})) are shown in Fig.~\ref{fig:inviscid_profs} for two of the previously considered datasets. In both cases, the present method for computing $U_I$ based on the Bernoulli equation leads to a significantly more accurate reconstruction of the velocity field $U$ outside the boundary layer. This implies that the present method, which leverages the wall-normal pressure profile, is more generally applicable than the two simplified models.

\section{Generalization of the local-reconstruction method for compressible flows} \label{app:compressible}

The local-reconstruction method presented in section~\ref{sec:new_method} can be readily generalized for compressible flows by replacing the stagnation pressure $P_o$ with the stagnation enthalpy $h_o$. Based on the assumption that the freestream flow is inviscid and non-heat-conducting outside the boundary layer, the stagnation enthalpy $h_o$ is approximately constant, i.e.
\begin{equation}
    h_o = h + \frac{U_m^2}{2} = const,
\end{equation}
where $h$ is the sensible enthalpy. Similar to the definition in incompressible flows, the locally reconstructed invicid solution is $U_I^2 = U_m^2 - V^2$. Then, $U_I[y]$ can be written in terms of $V[y]$, $h[y]$, and the reference stagnation enthalpy $h_{0,ref}$ as 
\begin{equation} \label{eq:general_h}
    U_I^2[y] = 2(h_{0,ref}-h[y]) - V^2[y].
\end{equation}
For a calorically perfect gas $h = C_p T$, where $C_p$ is a constant specific heat, the expression
\begin{equation} 
    U_I[y]^2 = \frac{2 \gamma}{\gamma - 1} \left(\frac{P_{ref}}{\rho_{ref}} - \frac{P[y]}{\rho[y]} \right) + U_{ref}^2 + V_{ref}^2 - V^2[y]
\end{equation}
holds.
Since the stagnation enthalpy is constant across shock waves, the present method can be deployed in flows containing shock waves, implying its potential extension to complex high-Mach flows \cite{fu2021shock}.


\bibliography{wall_model.bib,PG_NEq_LOW.bib,incomp_database.bib,incomp_wall.bib,del99.bib,comp_wall.bib}

\end{document}